\documentclass{cpbtex}
\usepackage{lineno,hyperref,amssymb,amsmath,epsfig,mathrsfs,subfigure}
\begin{document}
\begin{CJK*}{GBK}{song}

\title{On multi-soliton solutions to a generalized inhomogeneous nonlinear Schr\"odinger equation for the Heisenberg ferromagnetic spin chain\thanks{This work is supported by the National Natural Science Foundation of China (Grant No.~61775126).}}

% Title should be concise; avoid abbreviations if possible; and not begin with `A', `An', `The', or `Study on'.

\author{Zhou-Zheng Kang\thanks{Corresponding author. E-mail:~zhzhkang@126.com}, \ Rong-Cao Yang\\
{\small School of Physics $\&$ Electronic Engineering, Shanxi University, Taiyuan 030006, China}\\  % The line break was forced via \\
%$^{2}${\small Department of Mathematics, Shanghai University, Shanghai 200444, China} % The line break was forced via \\
}   % The line break was forced via \\

% 1. For Chinese authors, the name in Chinese characters should also be given. For example, Gang Liu(Áõ¸Õ), Xiao-Ming Li(ÀîÏþÃ÷)
% 2. Please ensure that every author approves the submission of the manuscript
% 3. Abbreviations should not be used in the affiliations

\date{}
\maketitle

\begin{abstract}
A generalized inhomogeneous higher-order nonlinear Schr\"odinger (GIHNLS) equation for the Heisenberg ferromagnetic spin chain system in (1+1)-dimensions under zero boundary condition at infinity is taken into account. The spectral analysis is first performed to generate a related matrix Riemann-Hilbert problem on the real axis. Then, through solving the resulting matrix Riemann-Hilbert problem by taking the jump matrix to be the identity matrix, the general bright multi-soliton solutions to the GIHNLS equation are attained. Furthermore, the one-, two-, and three-soliton solutions are written out and analyzed by figures.
\end{abstract}

\textbf{Keywords:} higher-order nonlinear Schr\"odinger equation; Riemann-Hilbert problem; soliton solutions

%\textbf{PACS:} no more than four \href{http://cpb.iphy.ac.cn/EN/column/item208.shtml}{PACS codes} should be provided

\section{Introduction}
Solitons are stable, nonlinear pulses which show a fine balance between nonlinearity and dispersion. They often arise from some real physical phenomena described by integrable nonlinear partial differential equations (NLPDEs) modelling shallow water waves, nonlinear optics, electrical network pulses and many other applications in mathematical physics [1-3].
Both theoretical and experimental investigations [4-6] have been made on solitons.
The derivation of abundant soliton solutions [7,8] to NLPDEs has been closely concerned by scholars from mathematics and physics, and a variety of approaches and their extentions have been established and applicable to NLPDEs up to present, such as the Hirota's bilinear method [9,10], the Darboux transformation [11,12], the Riemann-Hilbert method [13-15], and the Lie symmetry method [16,17].
In the past years, a considerable literature has grown up around the applications of the Riemann-Hilbert technique to solve integrable NLPDEs with zero or nonzero boundary condition, some of which include the coupled NLS equation [18], the Kundu-Eckhaus equation [19], the six-component fourth-order AKNS system [20], the multicomponent mKdV system [21], the $N$-coupled Hirota equation [22], and the fifth-order NLS equation [23].

In this paper, we focus on a generalized inhomogeneous higher-order nonlinear Schr\"odinger (GIHNLS) equation for the Heisenberg ferromagnetic spin system [26] in (1+1)-dimensions
\begin{equation}
iu_{t}+\epsilon u_{{{\it xxxx}}}+8\epsilon\left|u\right|^{2}u_{xx}+2\epsilon {u}^{2}
u_{xx}^{*}+4\epsilon u_{x}u_{x}^{*}u+6\epsilon u^{*}u_{x}^{2}+6\epsilon \left|u\right|^{4}u+(\frac{1}{2}-3\epsilon)u_{xx}+(1-6\epsilon){u}^{2}u^{*}-ihu_{x}=0,
\end{equation}
where $u$ denotes the complex function of the scaled spatial variable $x$ and temporal variable $t$, the real number $\epsilon$ is a perturbation parameter, the real number $h$ stands for the inhomogeneities in the medium [24,25], and the asterisk and subscripts mean the complex conjugation and partial derivatives, respectively. Equation (1) is an integrable model. When $h=0$, Eq. (1) reduces to the fourth-order NLS equation, which governs the Davydov solitons in the alpha helical protein with higher-order effects [27].
In the past, many studies have been conducted on Eq. (1). The Lax pair [24] was first presented. The gauge transformation was used to construct soliton solutions [26].
The generalized Darboux technique was applied to generate some higher-order rogue wave solutions [28].
In a follow-up study, some solutions were computed by Hirota's bilinear method, and infinitely many conservation laws were derived based upon the AKNS system [29].

The rest of the paper is arranged as follows. In Section 2, we formulate a matrix Riemann-Hilbert problem by carrying out the spectral analysis and obtain the reconstruction formula of potential. In Section 3,
we gain soliton solutions from a specific Riemann-Hilbert problem on the real axis, in which the jump matrix is taken as the identity matrix. The final section is a brief conclusion.

\section{Matrix Riemann-Hilbert problem}
What we intend to describe in this section is a matrix Riemann-Hilbert problem.
We start by considering the Lax pair [28] for Eq. (1)
\begin{align}
&{{\phi}_{x}}=U\phi,\\
&{{\phi}_{t}}=V\phi,
\end{align}
where ${\phi}={({\phi}_{1},{\phi}_{2})^\textrm{T}}$ is the spectral function, the symbol T stands for the vector transpose, and $\lambda\in \mathbb{C}$ is a spectral parameter. And
$$
U=\left(
    \begin{matrix}
      -i\lambda & u \\
      -u^{*} & i\lambda \\
    \end{matrix}
  \right),\quad
V=\left(
    \begin{matrix}
      V_{11} & V_{12} \\
      -V_{21} & -V_{11} \\
    \end{matrix}
  \right),$$
\begin{align*}
&V_{11}=2\epsilon\lambda u_{x}u^{*}-i\epsilon u_{x}u_{x}^{*}+i\epsilon u_{xx}u^{*}-4i\epsilon\lambda^{2}uu^{*}-2\epsilon\lambda uu_{x}^{*}+3i\epsilon u^{2}(u^{*})^{2}-3i\epsilon uu^{*}+i\epsilon uu_{xx}^{*}+\frac{1}{2}iuu^{*}-ih\lambda\\
&\quad\quad\ -i\lambda^{2}+8i\epsilon\lambda^{4}+6i\epsilon\lambda^{2},\\
&V_{12}=6i\epsilon uu_{x}u^{*}+4\epsilon\lambda u^{2}u^{*}+hu-4i\epsilon\lambda^{2}u_{x}+2\epsilon\lambda u_{xx}-3i\epsilon u_{x}+i\epsilon u_{xxx}+\frac{1}{2}iu_{x}+\lambda u-8\epsilon\lambda^{3}u-6\epsilon\lambda u,\\
&V_{21}=hu^{*}+\lambda u^{*}-8\epsilon\lambda^{3}u^{*}+4i\epsilon\lambda^{2}u_{x}^{*}+4\epsilon\lambda u(u^{*})^{2}-6\epsilon\lambda u^{*}+2\epsilon\lambda u_{xx}^{*}+3i\epsilon u_{x}^{*}-6i\epsilon uu^{*}u_{x}^{*}-i\epsilon u_{xxx}^{*}-\frac{1}{2}iu_{x}^{*}.
\end{align*}

Equivalently, the Lax pair (2) and (3) reads
\begin{align}
{\phi_{x}}&=(-i\lambda\Lambda+Q)\phi,\\
{\phi_{t}}&=\big((8i\epsilon\lambda^{4}+6i\epsilon\lambda^{2}-i\lambda^{2}-ih\lambda)\Lambda+V_{1}\big)\phi,
\end{align}
in which $\Lambda=\text{diag}(1,-1)$ and
$$
Q=\left(
    \begin{matrix}
     0 & u \\
     -u^{*} & 0 \\
    \end{matrix}
  \right),\quad
V_{1}=\left(
    \begin{matrix}
      \tilde{V}_{11} & \tilde{V}_{12} \\
      -\tilde{V}_{21} & -\tilde{V}_{11} \\
    \end{matrix}
  \right),$$
\begin{align*}
&\tilde{V}_{11}=2\epsilon\lambda u_{x}u^{*}-i\epsilon u_{x}u_{x}^{*}+i\epsilon u_{xx}u^{*}-4i\epsilon\lambda^{2}uu^{*}-2\epsilon\lambda uu_{x}^{*}+3i\epsilon u^{2}(u^{*})^{2}-3i\epsilon uu^{*}+i\epsilon uu_{xx}^{*}+\frac{1}{2}iuu^{*}\\
&\tilde{V}_{12}=6i\epsilon uu_{x}u^{*}+4\epsilon\lambda u^{2}u^{*}+hu-4i\epsilon\lambda^{2}u_{x}+2\epsilon\lambda u_{xx}-3i\epsilon u_{x}+i\epsilon u_{xxx}+\frac{1}{2}iu_{x}+\lambda u-8\epsilon\lambda^{3}u-6\epsilon\lambda u,\\
&\tilde{V}_{21}=hu^{*}+\lambda u^{*}-8\epsilon\lambda^{3}u^{*}+4i\epsilon\lambda^{2}u_{x}^{*}+4\epsilon\lambda u(u^{*})^{2}-6\epsilon\lambda u^{*}+2\epsilon\lambda u_{xx}^{*}+3i\epsilon u_{x}^{*}-6i\epsilon uu^{*}u_{x}^{*}-i\epsilon u_{xxx}^{*}-\frac{1}{2}iu_{x}^{*}.
\end{align*}

In our analysis, we suppose the potential $u$ to be vanished rapidly
at infinity. It is evident to see from (4) and (5) that $\phi \sim{{{e}}^{-i\lambda\Lambda x+(8i\epsilon\lambda^{4}+6i\epsilon\lambda^{2}-i\lambda^{2}-ih\lambda)\Lambda t}}.$
Thus we introduce the transformation
$$
\phi=\psi{{{e}}^{-i\lambda\Lambda x+(8i\epsilon\lambda^{4}+6i\epsilon\lambda^{2}-i\lambda^{2}-ih\lambda)\Lambda t}},
$$
which enable us to convert the Lax pair (4) and (5) into
\begin{align}
&{\psi_{{x}}}=-i\lambda[\Lambda,\psi]+Q\psi,\\
&{\psi_{t}}=(8i\epsilon\lambda^{4}+6i\epsilon\lambda^{2}-i\lambda^{2}-ih\lambda)[\Lambda,\psi]+V_{1}\psi,
\end{align}
where the square brackets denote the usual matrix commutator, namely, $[\Lambda,\psi]=\Lambda\psi-\psi\Lambda$.

In what follows, the spectral problem (6) will be analyzed,
and $t$ will be treated as a constant. We represent the matrix Jost solutions ${{\psi}_{\pm}}(x,\lambda)$ as
\begin{equation}
{{\psi}_{\pm}}(x,\lambda)=({{[{\psi_{\pm}}]_{1}}},{{[{\psi_{\pm}}]_{2}}})(x,\lambda),
\end{equation}
with the boundary conditions
\begin{equation}
{\psi_{\pm}}(x,\lambda)\to \mathbf{I}_{2},\quad x\to \pm\infty .
\end{equation}
The above subscripts in $\psi$ refer to which end of the $x$-axis the boundary conditions are required for, and
$\mathbf{I}_{2}$ is the identity matrix of size 2. Using the boundary conditions (9), one obtains Volterra-type integral equations
\begin{align}
&{\psi_{-}}(x,\lambda)=\mathbf{I}_{2}+\int_{-\infty }^{x}{{{{e}}^{-i\lambda \Lambda (x-z)}}Q(z){\psi_{-}}(z,\lambda){{{e}}^{i\lambda \Lambda (x-z)}}{d}z},\\
&{\psi_{+}}(x,\lambda)=\mathbf{I}_{2}-\int_{x}^{+\infty }{{{{e}}^{-i\lambda \Lambda (x-z)}}Q(z){\psi_{+}}(z,\lambda){{{e}}^{i\lambda \Lambda (x-z)}}{d}z}.
\end{align}
From (10) and (11), we find that ${{[{\psi_{+}}]_{1}}}$ and ${{[{\psi_{-}}]_{2}}}$ are analytic for $\lambda\in{\mathbb{C}_{-}}$ and continuous for
$\lambda\in {\mathbb{C}_{-}}\cup \mathbb{R}$, while ${{[{\psi_{-}}]_{1}}}$ and ${{[{\psi_{+}}]_{2}}}$ are analytic for $\lambda\in{\mathbb{C}_{+}}$ and continuous for $\lambda\in {\mathbb{C}_{+}}\cup \mathbb{R}$, where ${\mathbb{C}_{-}}$
and ${\mathbb{C}_{+}}$ are the lower and upper half $\lambda$-plane.
Applying the Abel's identity, we reveal that $\det{\psi_{\pm}}$ are independent
of $x$, since $\text{tr}Q=0$. Evaluating $\det{\psi_{-}}$ at $x=-\infty$ and $\det{\psi_{+}}$ at $x=+\infty$, we have $\det{\psi_{\pm}}=1$ for $\forall x$ and
$\lambda \in \mathbb{R}$. Due to matrix solutions of (4), ${\psi_{-}}{{{e}}^{-i\lambda \Lambda x}}$ and ${\psi_{+}}{{{e}}^{-i\lambda\Lambda x}}$ are linearly associated by the scattering matrix $S(\lambda)$
\begin{equation}
{{\psi}_{-}}{{{e}}^{-i\lambda \Lambda x}}={\psi_{+}}{{{e}}^{-i\lambda \Lambda x}}S(\lambda),\quad S(\lambda)=\left(                                                                                      \begin{matrix}
                                                                                         s_{11} & s_{12} \\
                                                                                         s_{21} & s_{22} \\
                                                                                       \end{matrix}
                                                                                     \right),
\quad \lambda \in \mathbb{R}.
\end{equation}
We notice that $\det{S(\lambda)}=1$ due to $\det{\psi_{\pm}(x,\lambda)}=1$.

A matrix Riemann-Hilbert problem is associated with two matrix analytic functions. In view of the analytic properties of $\psi_{\pm}$, the analytic function in ${\mathbb{C}_{+}}$ is given by
\begin{equation}
{{P}_{1}}(x,\lambda)=({{[{\psi_{-}}]_{1}}},{{[{\psi_{+}}]_{2}}})(x,\lambda)=\psi_{-}A_{1}+\psi_{+}A_{2},
\end{equation}
in which
\begin{equation}
A_{1}=\text{diag}(1,0),\quad A_{2}=\text{diag}(0,1).
\end{equation}

%Now, we examine the large-$\lambda$ asymptotic behavior of ${{P}_{1}}$.
Because ${{P}_{1}}$ solves (6),
we make an asymptotic expansion for ${{P}_{1}}$ at large-$\lambda$
\begin{equation*}
{{P}_{1}}=P_{1}^{(0)}+\frac{P_{1}^{(1)}}{\lambda}+\frac{P_{1}^{(2)}}{\lambda^{2}}+O\left(\frac{1}{\lambda^{3}}\right),\quad \lambda \to \infty ,
\end{equation*}
and substitute the asymptotic expansion into (6). Comparing the coefficients of the same
powers of $\lambda$ yields
$$O(1):P_{1x}^{(0)}=-i\big[\Lambda,P_{1}^{(1)}\big]+QP_{1}^{(0)};\quad O(\lambda): -i\big[\Lambda,P_{1}^{(0)}\big]=0.$$
Thus, we see that $P_{1}^{(0)}=\mathbf{I}_{2}$, namely,
${{P}_{1}}\to \mathbf{I}_{2}$ as $\lambda \in {\mathbb{C}_{+}}\to \infty.$

For construction of a matrix Riemann-Hilbert problem, we still need the analytic counterpart of $P_{2}$ in ${\mathbb{C}_{-}}$. Consider the adjoint equation of (6)
\begin{equation}
{{\kappa}_{x}}=-i\lambda[\Lambda,\kappa]-\kappa Q.
\end{equation}
One can verify that the inverse matrices
\begin{equation}
\psi_{\pm}^{-1}(x,\lambda)=\left( \begin{matrix}
   {[\psi_{\pm}^{-1}]^{1}}  \\
   {[\psi_{\pm}^{-1}]^{2}}  \\
\end{matrix} \right)(x,\lambda)
\end{equation}
solve (15).
Here $[\psi_{\pm}^{-1}]^{j}(j=1,2)$ signify the $j$-th row of $\psi_{\pm}^{-1}$,
and follow the boundary conditions $\psi_{\pm}^{-1}(x,\lambda)\rightarrow \mathbf{I}_{2}$ as $x\rightarrow\pm\infty.$ From (12), it follows immediately that
\begin{equation}
\psi_{-}^{-1}={{{e}}^{-i\lambda \Lambda x}}S^{-1}(\lambda){{{e}}^{i\lambda \Lambda x}}\psi_{+}^{-1},\quad \lambda \in \mathbb{R},
\end{equation}
where $S^{-1}(\lambda)={{({{r}_{lk}})}_{2\times 2}}$. Thus, the analytic function $P_{2}$ in ${\mathbb{C}_{-}}$ is expressed as
\begin{equation}
{{P}_{2}}(x,\lambda)=\left( \begin{matrix}
   {[\psi_{-}^{-1}]^{1}}  \\
   {[\psi_{+}^{-1}]^{2}}  \\
\end{matrix} \right)(x,\lambda)=A_{1}\psi_{-}^{-1}+A_{2}\psi_{+}^{-1},
\end{equation}
where $A_{1}$ and $A_{2}$ are given by (14).
One can find that the asymptotic behavior of $P_{2}$ turns out to be ${{P}_{2}}\to \mathbf{I}_{2}$ as $\lambda\to \infty .$

Inserting ${{\psi}_{\pm}}(x,\lambda)$ into (12) gives
$$
{{[{\psi_{-}}]_{1}}}={{s}_{11}}{{[{\psi_{+}}]_{1}}}+{{s}_{21}}{{{e}}^{2i\lambda x}}{{[{\psi_{+}}]_{2}}}.
$$

The $\psi_{\pm}^{-1}(x,\lambda)$  are then substituted into (17) yielding
$$
{[\psi_{-}^{-1}]^{1}}={{r}_{11}}{[\psi_{+}^{-1}]^{1}}+{{r}_{12}}{{{e}}^{-2i\lambda x}}{[\psi_{+}^{-1}]^{2}}.
$$

Thus, $P_{1}$ and $P_{2}$ can be represented as
$${{P}_{1}}=({{[{\psi_{-}}]_{1}}},{{[{\psi_{+}}]_{2}}})=({{[{\psi_{+}}]_{1}}},{{[{\psi_{+}}]_{2}}})\left( \begin{matrix}
   {{s}_{11}} & 0\\
   {{s}_{21}}{{{e}}^{2i\lambda x}} & 1\\
\end{matrix} \right),\quad
{{P}_{2}}=\left(\begin{matrix}
   {[\psi_{-}^{-1}]^{1}}  \\
   {[\psi_{+}^{-1}]^{2}}  \\
\end{matrix}\right)=\left(\begin{matrix}
   {{r}_{11}} & {{r}_{12}}{{{e}}^{-2i\lambda x}}  \\
    0 & 1\\
\end{matrix}\right)\left(\begin{matrix}
   {[\psi_{+}^{-1}]^{1}}  \\
   {[\psi_{+}^{-1}]^{2}}  \\
\end{matrix}\right).
$$

Having presented two matrix functions $P_{1}$ and $P_{2}$ which are analytic in ${\mathbb{C}_{+}}$ and ${\mathbb{C}_{-}}$, respectively,
a matrix Riemann-Hilbert problem on the real axis can be formed below
\begin{equation}
{{P}^{-}}(x,\lambda){{P}^{+}}(x,\lambda)=G(x,\lambda)=\left(\begin{matrix}
   1 & {{r}_{12}}{{{e}}^{-2i\lambda x}}  \\
   {{s}_{21}}{{{e}}^{2i\lambda x}} & 1  \\
\end{matrix}\right),\quad \lambda\in \mathbb{R},
\end{equation}
in which we have denoted that ${P_{1}}\rightarrow{P^{+}}$ as $\lambda\in {\mathbb{C}_{+}}\rightarrow\mathbb{R}$ and ${P_{2}}\rightarrow{P^{-}}$ as $\lambda\in {\mathbb{C}_{-}}\rightarrow\mathbb{R}$.
And the canonical normalization conditions are given by
$$
{{P}_{1}}(x,\lambda)\to \mathbf{I}_{2},\quad \lambda \in {\mathbb{C}_{+}}\to \infty;\quad
{{P}_{2}}(x,\lambda)\to \mathbf{I}_{2},\quad \lambda \in {\mathbb{C}_{-}}\to \infty.
$$

Next, we are going to present the reconstruction formula of the potential. Since $P_{1}(x,\lambda)$ solves (6), expanding
$P_{1}(x,\lambda)$ at large-$\lambda$ as
$$
{{P}_{1}}=\mathbf{I}_{2}+\frac{P_{1}^{(1)}}{\lambda}+\frac{P_{1}^{(2)}}{\lambda^{2}}+O\left(\frac{1}{\lambda^{3}}\right),\quad \lambda\to \infty,
$$
and inserting this expansion into (6), we see that
$$
Q=i\big[\Lambda ,P_{1}^{(1)}\big]=\left(\begin{matrix}
   0 & 2i\big(P_{1}^{(1)}\big)_{12}  \\
   -2i\big(P_{1}^{(1)}\big)_{21} & 0  \\
\end{matrix} \right)\Longrightarrow u=2i{{\big(P_{1}^{(1)}\big)_{12}}},
$$
where ${{\big(P_{1}^{(1)}\big)_{12}}}$ is the (1,2)-entry of $P_{1}^{(1)}$. By now, we have achieved the reconstruction for the potential.

\section{Soliton solutions}
For calculation of soliton solutions to Eq. (1),
we make an assumption that $\det {{P}_{1}}(\lambda)$ and $\det {{P}_{2}}(\lambda)$ can be zeros at certain discrete locations in analytic domains. Based on $\det{\psi_{\pm}}=1$, (13) and (18) as well as the scattering relation (12), we reveal that
$
\det {{P}_{1}}(\lambda)={s_{11}}(\lambda)$ and $\det {{P}_{2}}(\lambda)={r_{11}}(\lambda).
$
That is to say, $\det {{P}_{1}}(\lambda)$ and $\det {{P}_{2}}(\lambda)$ have the same zeros as ${s}_{11}(\lambda)$ and ${r}_{11}(\lambda)$. %and ${{r}_{22}}={{s}_{11}}$.
We now need the locations of zeros. Notice that the potential matrix $Q$ satisfies the anti-Hermitian property $Q^{\dagger }=-Q$, where $\dagger$ means the matrix Hermitian. Taking advantage of this property in $Q$, one has
\begin{equation}
\psi_{\pm }^{\dagger }(x,{\lambda}^{*})=\psi_{\pm }^{-1}(x,\lambda).
\end{equation}
After taking the Hermitian to (13) and considering (18), we find that
\begin{equation}
P_{1}^{\dagger }({{\lambda}^{*}})={{P}_{2}}(\lambda),\quad \lambda \in {\mathbb{C}_{-}},
\end{equation}
and
$${{S}^{\dagger }}(\lambda^{*})={{S}^{-1}}(\lambda)\Longrightarrow
s_{11}^{*}({{\lambda}^{*}})={{r}_{11}}(\lambda).$$
From this, it is found that each zero ${\lambda_{j}}$ of $\det {{P}_{1}}$ produces each zero $\lambda_{j}^{*}$ of $\det{{P}_{2}}$. Let $N$ be a free natural number. Generally, we
assume that $\det {{P}_{1}}$ and $\det{{P}_{2}}$ have some simple zeros at ${{\lambda}_{j}}\in{\mathbb{C}_{+}}$ and ${{\hat{\lambda}}_{j}}=\lambda_{j}^{*}\in{\mathbb{C}_{-}}$, respectively. For this case, each of the kernel of ${{P}_{1}}({{\lambda}_{j}})$ and ${{P}_{2}}({{\hat{\lambda}}_{j}})$ contains a single basis column vector ${{\nu}_{j}}$ or row vector ${{\hat{\nu}}_{j}}$:
\begin{align}
&{{P}_{1}}({{\lambda}_{j}}){{\nu}_{j}}=0,\\
&{{\hat{\nu}}_{j}}{{P}_{2}}({{\hat{\lambda}}_{j}})=0,
\end{align}
By taking the Hermitian to (22) and utilizing (21), we get
\begin{equation}
{{\hat{\nu}}_{j}}=\nu_{j}^{\dagger },\quad 1\le j\le N.
\end{equation}
Then computing $x$-derivative and $t$-derivative in (22) respectively, and using (6) and (7) yields
$${{P}_{1}}({{\lambda}_{j}})\left( \frac{\partial {{\nu}_{j}}}{\partial x}+i{\lambda_{j}}\Lambda {{\nu}_{j}} \right)=0,\quad
{{P}_{1}}({{\lambda}_{j}})\left( \frac{\partial {{\nu}_{j}}}{\partial t}-(8i\epsilon\lambda_{j}^{4}+6i\epsilon\lambda_{j}^{2}-i\lambda_{j}^{2}-ih\lambda_{j})\Lambda{{\nu}_{j}} \right)=0.$$
Therefore, we get
$$
{\nu_{j}}={{{e}}^{\left(-i{\lambda_{j}}x+(8i\epsilon\lambda_{j}^{4}+6i\epsilon\lambda_{j}^{2}-i\lambda_{j}^{2}-ih\lambda_{j})t\right)\Lambda }}{\nu_{j0}}.
$$
In view of the relation (24), we see that
$${{\hat{\nu}}_{j}}=\nu_{j0}^{\dagger }{{{e}}^{\left(i{\lambda_{j}^{*}}x-(8i\epsilon{\lambda_{j}^{*}}^{4}+6i\epsilon{\lambda_{j}^{*}}^{2}-i{\lambda_{j}^{*}}^{2}-ih\lambda_{j}^{*})t\right)\Lambda}},$$
where $\nu_{j0}$ and $\nu_{j0}^{\dagger }$ are constants.

For presenting soliton solutions, we consider the reflectionless case, namely, $G(x,\lambda)=\mathbf{I}_{2}$.
This resulting special Riemann-Hilbert problem [30] possesses the solutions
\begin{equation} {{P}_{1}}(\lambda)=\mathbf{I}_{2}-\sum\limits_{k=1}^{N}{\sum\limits_{j=1}^{N}{\frac{{\nu_{k}}{{{\hat{\nu}}}_{j}}{{\big({{M}^{-1}}\big)_{kj}}}}{\lambda -{{{\hat{\lambda}}}_{j}}}}}, \quad {{P}_{2}}(\lambda)=\mathbf{I}_{2}+\sum\limits_{k=1}^{N}{\sum\limits_{j=1}^{N}{\frac{{\nu_{k}}{{{\hat{\nu}}}_{j}}{{\big({{M}^{-1}}\big)_{kj}}}}{\lambda-{{\lambda }_{k}}}}},
\end{equation}
where $M=({{m}_{kj}})_{N\times N}$ and
\begin{equation*}
{{m}_{kj}}=\frac{{\hat{\nu}_{k}}{{{{\nu}}}_{j}}}{{{\lambda}_{j}}-{{{\hat{\lambda}}}_{k}}},\quad 1\le k,j\le N.
\end{equation*}
From Eq. (25), we derive
\begin{equation*}
P_{1}^{(1)}=-\sum\limits_{k=1}^{N}{\sum\limits_{j=1}^{N}{{{\nu}_{k}}{{{\hat{\nu}}}_{j}}{{\big({{M}^{-1}}\big)_{kj}}}}}.
\end{equation*}

Combining the established results with ${{\nu}_{j0}}={({a_{j}},{b_{j}})^\textrm{T}}$ and
${{\vartheta}_{j}}=-i{\lambda_{j}}x+(8i\epsilon\lambda_{j}^{4}+6i\epsilon\lambda_{j}^{2}-i\lambda_{j}^{2}-ih\lambda_{j})t$, the general $N$-soliton solution to Eq. (1) can be written as
\begin{equation*}
u(x,t)=-2i\sum\limits_{k=1}^{N}{\sum\limits_{j=1}^{N}{a _{k}{b_{j}^{*}}{{{e}}^{{\vartheta}_{k}-{\vartheta_{j}^{*}}}}{{\big({{M}^{-1}}\big)_{kj}}}}},
\quad
{m}_{kj}=\frac{1}{{{\lambda}_{j}}-\lambda_{k}^{*}}{\big(a_{k}^{*}{a_{j}}{{{e}}^{\vartheta_{k}^{*}+{{\vartheta}_{j}}}}+b_{k}^{*}{ b_{j}}{{{e}}^{-\vartheta_{k}^{*}-{{\vartheta }_{j}}}}\big)}.
\end{equation*}

In what follows, we intend to discuss one-, two-, and three-soliton solutions graphically.

(i) For $N=1$, Eq. (1) possesses one-soliton solution
\begin{equation}
u(x,t)=\frac{4{{a}_{1}}b_{1}^{*}{\lambda_{12}}{{{e}}^{{{\vartheta}_{1}}-\vartheta_{1}^{*}}}}{{{\left| {a_{1}} \right|}^{2}}{{{e}}^{\vartheta_{1}^{*}+{{\vartheta}_{1}}}}+{{\left| {b_{1}} \right|}^{2}}{{{e}}^{-\vartheta_{1}^{*}-{{\vartheta}_{1}}}}},
\end{equation}
where we have assumed ${{\vartheta}_{1}}=-i{\lambda_{1}}x+(8i\epsilon\lambda_{1}^{4}+6i\epsilon\lambda_{1}^{2}-i\lambda_{1}^{2}-ih\lambda_{1})t$ and ${\lambda_{1}}={\lambda_{11}}+i{\lambda_{12}}$.
Further, upon assuming ${b_{1}}=1$ and ${{\left| {a_{1}} \right|}^{2}}={{{e}}^{2{{\xi }_{1}}}}$, then the solution (26) can be written as
\begin{equation}
u(x,t)=2{{a}_{1}}{\lambda_{12}}{{{e}}^{-{{\xi}_{1}}}}{{{e}}^{{\vartheta_{1}}-{{\vartheta}_{1}^{*}}}}\text{sech}(\vartheta_{1}^{*}+{{\vartheta}_{1}}+{{\xi }_{1}}),
\end{equation}
where
$\vartheta_{1}^{*}+{\vartheta}_{1}=2\lambda_{12}x+(64\epsilon\lambda_{11}{\lambda_{12}^{3}}-64\epsilon{\lambda_{11}^{3}}\lambda_{12}-24\epsilon \lambda_{11}\lambda_{12}+2h\lambda_{12}+4\lambda_{{11}}\lambda_{12})t
$ and $
\vartheta_{1}-{\vartheta}_{1}^{*}=-2i\lambda_{11}x-2i(-8\epsilon {\lambda_{11}^{4}}+(48
\epsilon {\lambda_{12}^{2}}-6\epsilon +1){\lambda_{11}^{2}}+h\lambda_{11}+( -8{\lambda_{12}^{4}}+6{\lambda_{12}^{2}})\epsilon-{\lambda_{12}^{2}})t.$
Equivalently, the solution (27) reads
\begin{equation}
u(x,t)=2{{a}_{1}}{\lambda_{12}}{{{e}}^{-{{\xi}_{1}}}}{{{e}}^{{\vartheta_{1}}-{{\vartheta}_{1}^{*}}}}\text{sech}\big[2\lambda_{12}x+\big(64\epsilon\lambda_{11}{\lambda_{12}^{3}}-64\epsilon{\lambda_{11}^{3}}\lambda_{12}-24\epsilon \lambda_{11}\lambda_{12}+2h\lambda_{12}+4\lambda_{{11}}\lambda_{12}\big)t+{{\xi}_{1}}\big].
\end{equation}

According to expression (28), we know that the amplitude function $|u(x,t)|$ has a sech profile. This soliton with peak amplitude $2|{{a}_{1}}|{\lambda_{12}}{{{e}}^{-{{\xi }_{1}}}}$ travels at velocity $-32\epsilon\lambda_{11}{\lambda_{12}^{2}}+32\epsilon{\lambda_{11}^{3}}+12\epsilon \lambda_{11}-h-2\lambda_{11}$ depending on both real and imaginary parts of the spectral parameter ${\lambda}_{1}$,
unlike the basic NLS equation. The phase relies on the spatial variable $x$ and temporal variable $t$ linearly. The soliton in Fig. 1 is formed for $a_{{1}}=1,\lambda_{1}=\frac{1}{6}+\frac{i}{2},\xi_{1}=0,\epsilon=1,h=1$. From Fig. 1(b), it is seen that the wave travels from right to left along the $x$-axis over time.
\begin{figure}
\begin{center}
\subfigure[]{\resizebox{0.31\hsize}{!}{\includegraphics*{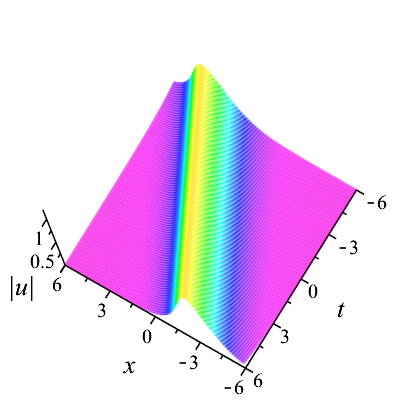}}}
\subfigure[]{\resizebox{0.31\hsize}{!}{\includegraphics*{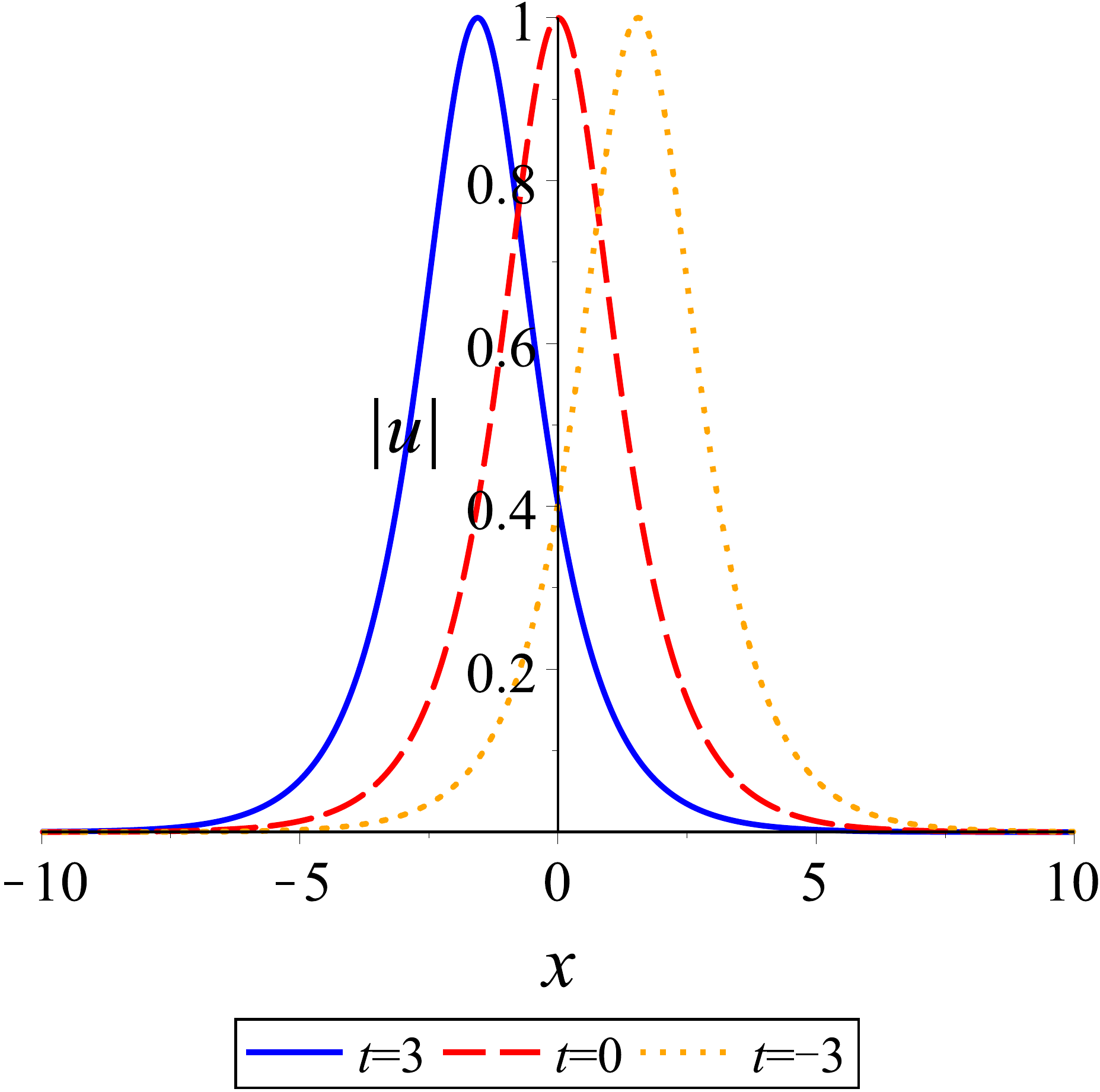}}}
\caption{Profiles of one-soliton solution (28) with $a_{{1}}=1,\lambda_{1}=\frac{1}{6}+\frac{i}{2},\xi_{1}=0,\epsilon=1,h=1$. (a) 3D plot; (b) $x$-curves.}
\end{center}
\end{figure}

(ii) For $N=2$, two-soliton solution is given by
\begin{equation}
u(x,t)=-\frac{2i\big(a _{1}{b_{1}^{*}}{{m}_{22}}{{{e}}^{{{\vartheta}_{1}}-\vartheta_{1}^{*}}}-a_{1}{b_{2}^{*}}{{m}_{12}}{{{e}}^{{{\vartheta }_{1}}-\vartheta_{2}^{*}}}-a _{2}{b_{1}^{*}}{{m}_{21}}{{{e}}^{{{\vartheta}_{2}}-\vartheta_{1}^{*}}}+a_{2}{{b }_{2}^{*}}{{m}_{11}}{{{e}}^{\vartheta_{2}-{{\vartheta}_{2}^{*}}}}\big)}{{{m}_{11}}{{m}_{22}}-{{m}_{12}}{{m}_{21}}},
\end{equation}
where
$$\begin{aligned}
&{{m}_{11}}=\frac{1}{{\lambda_{1}}-\lambda_{1}^{*}}{\big({{\left| {a_{1}} \right|}^{2}}{{{e}}^{\vartheta_{1}^{*}+{{\vartheta}_{1}}}}+{{\left| {b_{1}} \right|}^{2}}{{{e}}^{-\vartheta _{1}^{*}-{{\vartheta}_{1}}}}\big)},\quad
{{m}_{12}}=\frac{1}{{\lambda_{2}}-\lambda _{1}^{*}}{\big(a_{1}^{*}{a_{2}}{{{e}}^{\vartheta_{1}^{*}+{{\vartheta}_{2}}}}+b_{1}^{*}{b_{2}}{{{e}}^{-\vartheta_{1}^{*}-{{\vartheta }_{2}}}}\big)}, \\
&{{m}_{21}}=\frac{1}{{\lambda_{1}}-\lambda_{2}^{*}}{\big(a_{2}^{*}{a_{1}}{{{e}}^{\vartheta_{2}^{*}+{{\vartheta}_{1}}}}+b_{2}^{*}{b_{1}}{{{e}}^{-\vartheta_{2}^{*}-{{\vartheta }_{1}}}}\big)},\quad
{{m}_{22}}=\frac{1}{{\lambda_{2}}-\lambda_{2}^{*}}{\big({{\left| {a_{2}} \right|}^{2}}{{{e}}^{\vartheta_{2}^{*}+{{\vartheta}_{2}}}}+{{\left| {b_{2}} \right|}^{2}}{{{e}}^{-\vartheta _{2}^{*}-{{\vartheta}_{2}}}}\big)},
\end{aligned}$$
and
${{\vartheta}_{\iota}}=-i{\lambda_{\iota}}x+(8i\epsilon\lambda_{\iota}^{4}+6i\epsilon\lambda_{\iota}^{2}-i\lambda_{\iota}^{2}-ih\lambda_{\iota})t,\lambda_{\iota}={\lambda_{\iota 1}}+i{\lambda_{\iota 2}},\iota=1,2$.
Through assuming $a_{1}=a_{2},{b_{1}}={b_{2}}=1$, and ${{\left| {a_{1}}\right|}^{2}}={{{e}}^{2{{\xi }_{1}}}}$, the solution (29) reads
\begin{equation}
u(x,t)=-\frac{2i\big(a _{1}{{m}_{22}}{{{e}}^{{{\vartheta}_{1}}-\vartheta_{1}^{*}}}-a_{1}{{m}_{12}}{{{e}}^{{{\vartheta}_{1}}-\vartheta_{2}^{*}}}-a _{2}{{m}_{21}}{{{e}}^{{{\vartheta}_{2}}-\vartheta_{1}^{*}}}+a_{2}{{m}_{11}}{{{e}}^{\vartheta_{2}-{{\vartheta}_{2}^{*}}}}\big)}{{{m}_{11}}{{m}_{22}}-{{m}_{12}}{{m}_{21}}},
\end{equation}
in which
$$
\begin{aligned}
&{{m}_{11}}=-\frac{i{{{e}}^{{{\xi }_{1}}}}}{{{\lambda}_{12}}}\cosh(\vartheta_{1}^{*}+{{\vartheta}_{1}}+{{\xi }_{1}}),\quad {{m}_{12}}=\frac{2{{{e}}^{{{\xi }_{1}}}}}{({{\lambda}_{21}}-{{\lambda}_{11}})+i({{\lambda}_{12}}+{{\lambda}_{22}})}\cosh(\vartheta_{1}^{*}+{{\vartheta}_{2}}+{{\xi }_{1}}),\\
&{{m}_{22}}=-\frac{i{{{e}}^{{{\xi }_{1}}}}}{{\lambda_{22}}}\cosh(\vartheta_{2}^{*}+{{\vartheta}_{2}}+{{\xi }_{1}}),\quad {{m}_{21}}=\frac{2{{{e}}^{{{\xi }_{1}}}}}{({{\lambda}_{11}}-{{\lambda}_{21}})+i({{\lambda}_{12}}+{{\lambda}_{22}})}\cosh(\vartheta_{2}^{*}+{{\vartheta}_{1}}+{{\xi }_{1}}).
\end{aligned}
$$

In order to show interaction behaviors between two solitons, some graphs are plotted and two cases are under consideration here.

We first consider the case of two solitons traveling at different velocities.
In this case,
the solution parameters in (30) are first chosen as $a_{1}=1,a_{2}=1,\lambda_{1}=\frac{1}{10}+\frac{i}{3},
\lambda_{2}=\frac{1}{10}+\frac{i}{2},\xi_{1}=0,\epsilon =1,h=1$.
According to these values, some plots are made to shed light on the localization and dynamical behaviors.
Figure 2(a) shows the localized structure of this solution on $(x,t)$-plane clearly, which is a typical cross two bright solitons. It can be observed that the overtaking collision between the solitons takes place as depicted in Fig. 2(b), where two solitons with different velocities move together towards the same direction along the $x$-axis. The (taller) soliton with a larger amplitude travels much faster than the other (shorter) soliton with a smaller amplitude, and the taller soliton catches up with the shorter soliton over time. Both solitons then continue to proceed in the same direction. At the moment $t=0$, the amplitude value for two solitons reaches the maximum. Before and after the collision, their speeds and shapes are unchanged.
In other words, the overtaking is an elastic interaction.

In Fig. 3, we show the
head-on collision between two solitons with the parameters as
$a_{1}=1,a_{2}=1,\lambda_{1}=\frac{1}{10}+\frac{i}{2},
\lambda_{2}=\frac{1}{6}+\frac{i}{3},\xi_{1}=0,\epsilon =1,h=1$. The taller soliton crashes the shorter one in the opposite direction of the $x$-axis. After the collision,
their amplitudes, widths, speeds, and directions are same as those before only except phase shifts, see Fig. 3(b). Evidently, the head-on interaction of two solitons is also elastic.
%It is found that the two bright solitons keep their amplitudes, widths and directions invariant during propagation.
\begin{figure}
\begin{center}
\subfigure[]{\resizebox{0.31\hsize}{!}{\includegraphics*{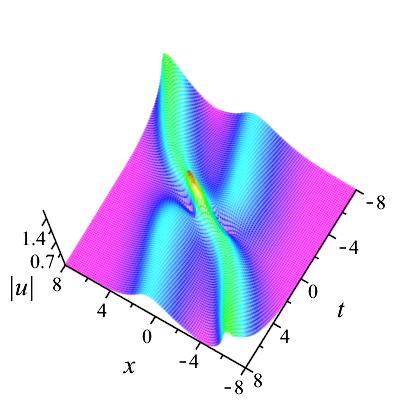}}}
\subfigure[]{\resizebox{0.31\hsize}{!}{\includegraphics*{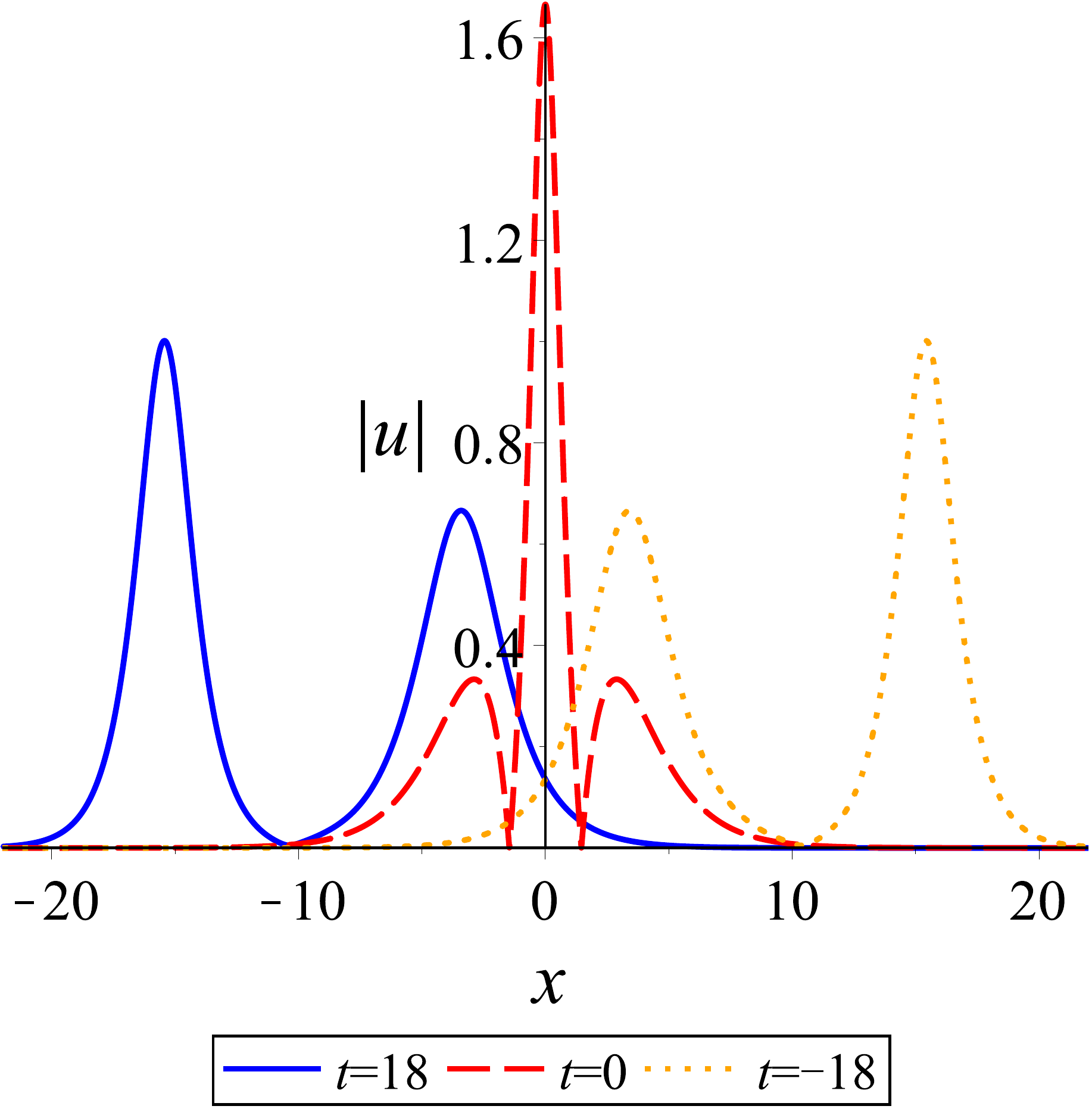}}}
\caption{Profiles of two-soliton solution (30) with $a_{1}=1,a_{2}=1,
\lambda_{1}=\frac{1}{10}+\frac{i}{3},\lambda_{2}=\frac{1}{10}+\frac{i}{2},\xi_{1}=0,\epsilon =1,h=1
$. (a) 3D plot; (b) $x$-curves.}
\end{center}
\end{figure}
\begin{figure}
\begin{center}
\subfigure[]{\resizebox{0.31\hsize}{!}{\includegraphics*{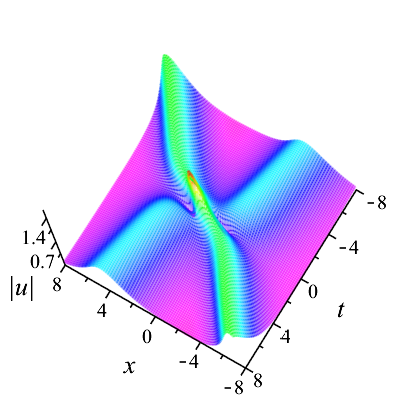}}}
\subfigure[]{\resizebox{0.31\hsize}{!}{\includegraphics*{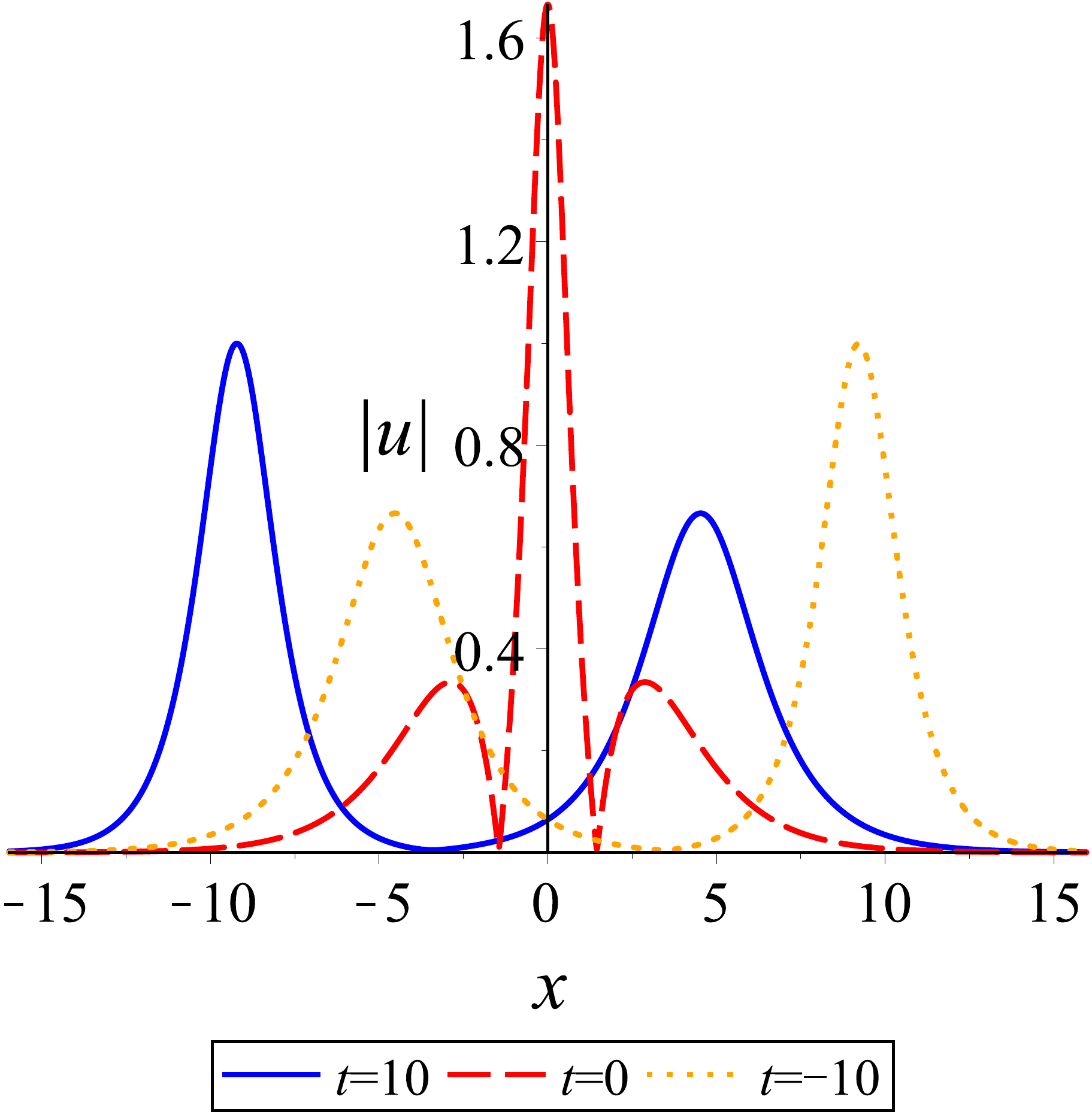}}}
\caption{Profiles of two-soliton solution (30) with $a_{1}=1,a_{2}=1,\lambda_{1}=\frac{1}{10}+\frac{i}{2},
\lambda_{2}=\frac{1}{6}+\frac{i}{3},\xi_{1}=0,\epsilon =1,h=1
$. (a) 3D plot; (b) $x$-curves.}
\end{center}
\end{figure}

With regard to the second case, we consider that two solitons travel at same speeds.
The solution parameters in (30) are specified as
$a_{1}=1,a_{2}=1,\lambda_{1}=\frac{i}{3},\lambda_{2}=\frac{i}{2},\xi_{1}=0,\epsilon =1,h=1$.
The bound state of two solitons is shown on $(x,t)$-plane in Fig. 4,
in which two solitons are localized spatially and keep together in propagation.
Indeed, this solution indicates the breather, namely, when two solitons propagate, the amplitude function is periodic in oscillation over time.
\begin{figure}
\begin{center}
\subfigure[]{\resizebox{0.31\hsize}{!}{\includegraphics*{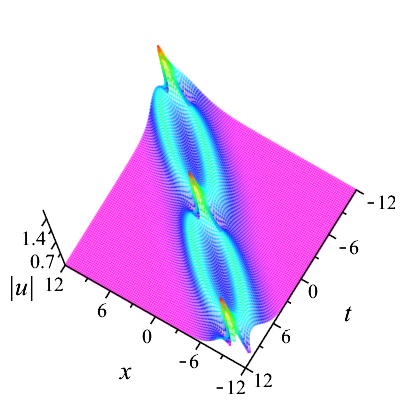}}}
\subfigure[]{\resizebox{0.31\hsize}{!}{\includegraphics*{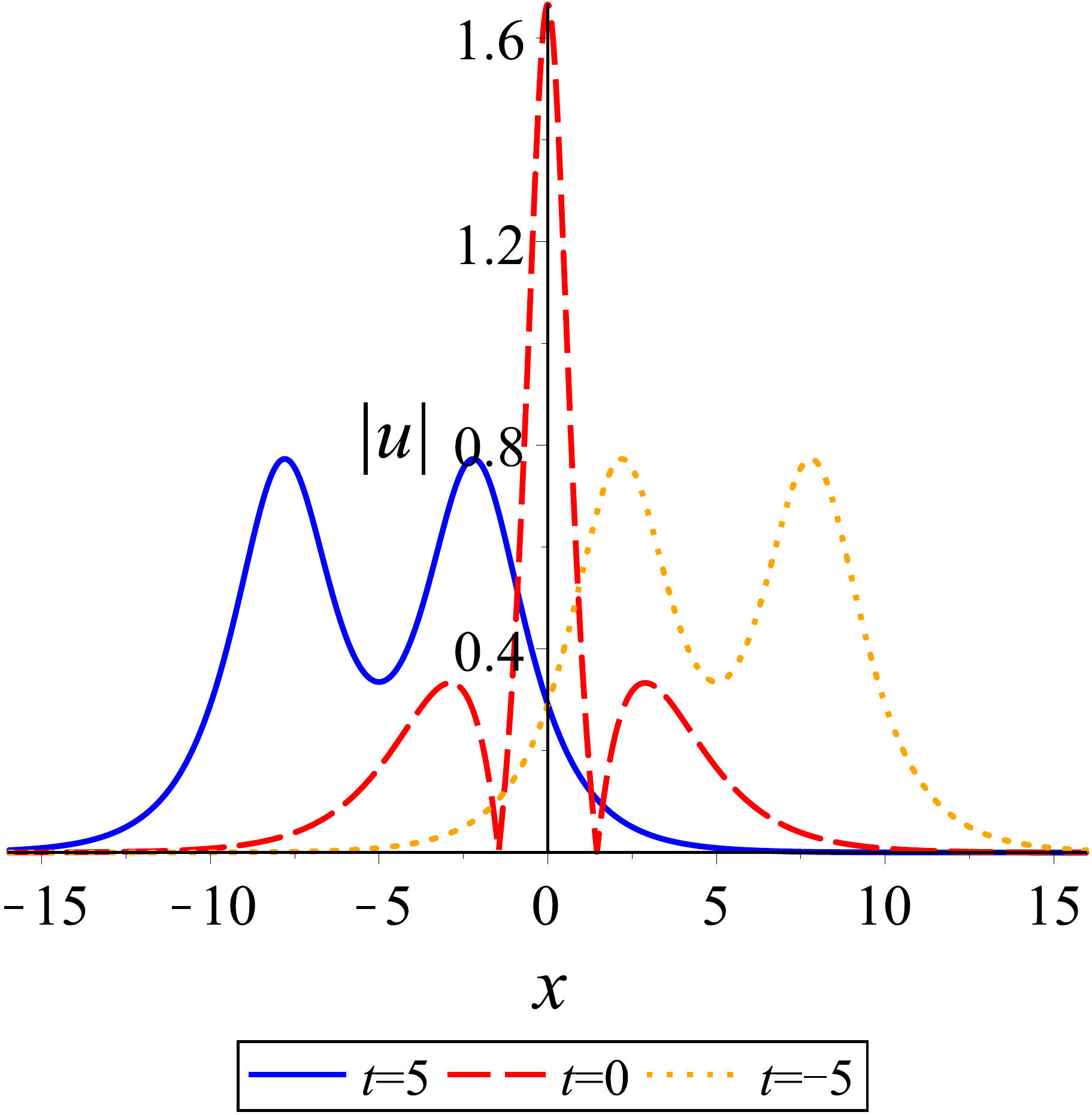}}}
\caption{Profiles of two-soliton solution (30) with $a_{1}=1,a_{2}=1,
\lambda_{1}=\frac{i}{3},\lambda_{2}=\frac{i}{2},\xi_{1}=0,\epsilon =1,h=1$. (a) 3D plot; (b) $x$-curves.}
\end{center}
\end{figure}

(iii) For $N=3$, Eq. (1) admits three-soliton solution
\begin{equation}
\begin{aligned}
u(x,t)=&-\frac{2i}{\varrho}\big[a_{1}{ b_{1}^{*}}{(m_{{22}}m_{{33}}-m_{{23}}m_{{32}})}{{{e}}^{{{\vartheta}_{1}}-\vartheta_{1}^{*}}}-a _{1}{b_{2}^{*}}{(m_{{12}}m_{{33}}-m_{{13}}m_{{32}})}{{{e}}^{{{\vartheta }_{1}}-\vartheta_{2}^{*}}}\\
&\quad\quad+a_{1}{b_{3}^{*}}{(m_{{12}}m_{{23}}-m_{{13}}m_{{22}})}{{{e}}^{{{\vartheta }_{1}}-\vartheta_{3}^{*}}}-a_{2}{b_{1}^{*}}{(m_{{21}}m_{{33}}-m_{{23}}m_{{31}})}{{{e}}^{{{\vartheta}_{2}}-\vartheta_{1}^{*}}}\\
&\quad\quad+a_{2}{ b_{2}^{*}}{(m_{{11}}m_{{33}}-m_{{13}}m_{{31}})}{{{e}}^{{{\vartheta}_{2}}-\vartheta_{2}^{*}}}-a _{2}{b_{3}^{*}}{(m_{{11}}m_{{23}}-m_{{13}}m_{{21}})}{{{e}}^{{{\vartheta}_{2}}-\vartheta_{3}^{*}}}\\
&\quad\quad+a_{3}{b_{1}^{*}}{(m_{{21}}m_{{32}}-m_{{22}}m_{{31}})}{{{e}}^{{{\vartheta}_{3}}-\vartheta_{1}^{*}}}-a _{3}{b_{2}^{*}}{(m_{{11}}m_{{32}}-m_{{12}}m_{{31}})}{{{e}}^{{{\vartheta }_{3}}-\vartheta_{2}^{*}}}\\
&\quad\quad+a_{3}{b_{3}^{*}}{(m_{{11}}m_{{22}}-m_{{12}}m_{{21}})}{{{e}}^{{{\vartheta }_{3}}-\vartheta_{3}^{*}}}\big],
\end{aligned}
\end{equation}
where $\varrho=m_{{11}}m_{{22}}m_{{33}}-m_{{11}}m_{{23}}m_{{32}}-m_{{12}}m_{{21}}m_{{33}}+m_{{12}}m_{{23}}m_{{31}}+m_{{13}}m_{{21}}m_{{32}}-m_{{13}}m_{{22}}m_{{31}},$
$$\begin{aligned}
&{{m}_{11}}=\frac{1}{{\lambda_{1}}-\lambda_{1}^{*}}{\big({{\left| {a_{1}} \right|}^{2}}{{{e}}^{\vartheta_{1}^{*}+{{\vartheta}_{1}}}}+{{\left| {b_{1}} \right|}^{2}}{{{e}}^{-\vartheta _{1}^{*}-{{\vartheta}_{1}}}}\big)},\quad
{{m}_{12}}=\frac{1}{{\lambda_{2}}-\lambda _{1}^{*}}{\big(a_{1}^{*}{a_{2}}{{{e}}^{\vartheta_{1}^{*}+{{\vartheta}_{2}}}}+b_{1}^{*}{b_{2}}{{{e}}^{-\vartheta_{1}^{*}-{{\vartheta }_{2}}}}\big)},\\
&{{m}_{13}}=\frac{1}{{\lambda_{3}}-\lambda _{1}^{*}}{\big(a_{1}^{*}{a_{3}}{{{e}}^{\vartheta_{1}^{*}+{{\vartheta}_{3}}}}+b_{1}^{*}{b_{3}}{{{e}}^{-\vartheta_{1}^{*}-{{\vartheta }_{3}}}}\big)},\quad
{{m}_{21}}=\frac{1}{{\lambda_{1}}-\lambda_{2}^{*}}{\big(a_{2}^{*}{a_{1}}{{{e}}^{\vartheta_{2}^{*}+{{\vartheta}_{1}}}}+b_{2}^{*}{b_{1}}{{{e}}^{-\vartheta_{2}^{*}-{{\vartheta }_{1}}}}\big)},\\
&{{m}_{22}}=\frac{1}{{\lambda_{2}}-\lambda_{2}^{*}}{\big({{\left| {a_{2}} \right|}^{2}}{{{e}}^{\vartheta_{2}^{*}+{{\vartheta}_{2}}}}+{{\left| {b_{2}} \right|}^{2}}{{{e}}^{-\vartheta _{2}^{*}-{{\vartheta}_{2}}}}\big)},\quad
{{m}_{23}}=\frac{1}{{\lambda_{3}}-\lambda _{2}^{*}}{\big(a_{2}^{*}{a_{3}}{{{e}}^{\vartheta_{2}^{*}+{{\vartheta}_{3}}}}+b_{2}^{*}{b_{3}}{{{e}}^{-\vartheta_{2}^{*}-{{\vartheta }_{3}}}}\big)},\\
&{{m}_{31}}=\frac{1}{{\lambda_{1}}-\lambda_{3}^{*}}{\big(a_{3}^{*}{a_{1}}{{{e}}^{\vartheta_{3}^{*}+{{\vartheta}_{1}}}}
+b_{3}^{*}{b_{1}}{{{e}}^{-\vartheta _{3}^{*}-{{\vartheta}_{1}}}}\big)},\quad
{{m}_{32}}=\frac{1}{{\lambda_{2}}-\lambda _{3}^{*}}{\big(a_{3}^{*}{a_{2}}{{{e}}^{\vartheta_{3}^{*}+{{\vartheta}_{2}}}}+b_{3}^{*}{b_{2}}{{{e}}^{-\vartheta_{3}^{*}-{{\vartheta }_{2}}}}\big)},\\
&{{m}_{33}}=\frac{1}{{\lambda_{3}}-\lambda _{3}^{*}}{\big({{\left| {a_{3}} \right|}^{2}}{{{e}}^{\vartheta_{3}^{*}+{{\vartheta}_{3}}}}+{{\left| {b_{3}} \right|}^{2}}{{{e}}^{-\vartheta_{3}^{*}-{{\vartheta }_{3}}}}\big)},
\end{aligned}$$
and ${{\vartheta}_{\iota}}=-i{\lambda_{\iota}}x+(8i\epsilon\lambda_{\iota}^{4}+6i\epsilon\lambda_{\iota}^{2}-i\lambda_{\iota}^{2}-ih\lambda_{\iota})t,\lambda_{\iota}={\lambda_{\iota 1}}+i{\lambda_{\iota 2}},\iota=1,2,3$.

Following the similar lines as our disscussion on two solitons, we now examine the dynamics among three solitons.
The parameter values in (31) are first given by  $a_{1}=1,a_{2}=1,a_{3}=1,b_{1}=1,b_{2}=1,b_{3}=1,\lambda_{1}=\frac{1}{10}+\frac{i}{2},
\lambda_{2}=\frac{1}{10}+\frac{2i}{3},\lambda_{3}=\frac{1}{10}+\frac{i}{3},\epsilon =1,h=1$. Based on these values, a special solution can be gained at once. And we can know the velocity relation for the three solitons $S_{1}\textless S_{2}\textless S_{3}$. Here we have denoted that the solitons from left to right in Fig. 5(a) are $S_{1},S_{2}$, and $S_{3}$.
Figure 5 presents an elastic overtaking process among three solitons moving together towards the negative direction of the $x$-axis. Ultimately as time evolves, $S_{2}$ overtakes $S_{1}$, and  $S_{3}$ overtakes $S_{1}$ and $S_{2}$.
%during which
When $t=0$, the peak amplitude is maximum.

Then, we take the parameters as $a_{1}=1,a_{2}=1,a_{3}=1,b_{1}=1,b_{2}=1,b_{3}=1,
\lambda_{1}=\frac{1}{10}+\frac{i}{2},\lambda_{2}=\frac{1}{6}+\frac{i}{3},\lambda_{3}=\frac{1}{8}+i,\epsilon=1,h=1.$
Denoting that the solitons from left to right in Fig. 6(a) are $s_{1},s_{2},$ and $s_{3}$ respectively, it is found in Fig. 6 that
$s_{1}$ moves towards the positive direction of the $x$-axis, which is opposite to the propagation direction of $s_{2}$ and $s_{3}$. As time goes on, $s_{1}$ collides head-on with $s_{2}$ and $s_{3}$, while $s_{3}$ overtakes $s_{2}$. After the interactions, the three solitons $s_{1},s_{2},$ and $s_{3}$ continue to move along their original directions. Both head-on and overtaking interactions in the process are elastic.
Additionally, the head-on interaction of two solitons in bound state with another soliton during propagation can be observed in Fig. 7. And Fig. 8 shows the evolution of bound state of three solitons.
\begin{figure}
\begin{center}
\subfigure[]{\resizebox{0.31\hsize}{!}{\includegraphics*{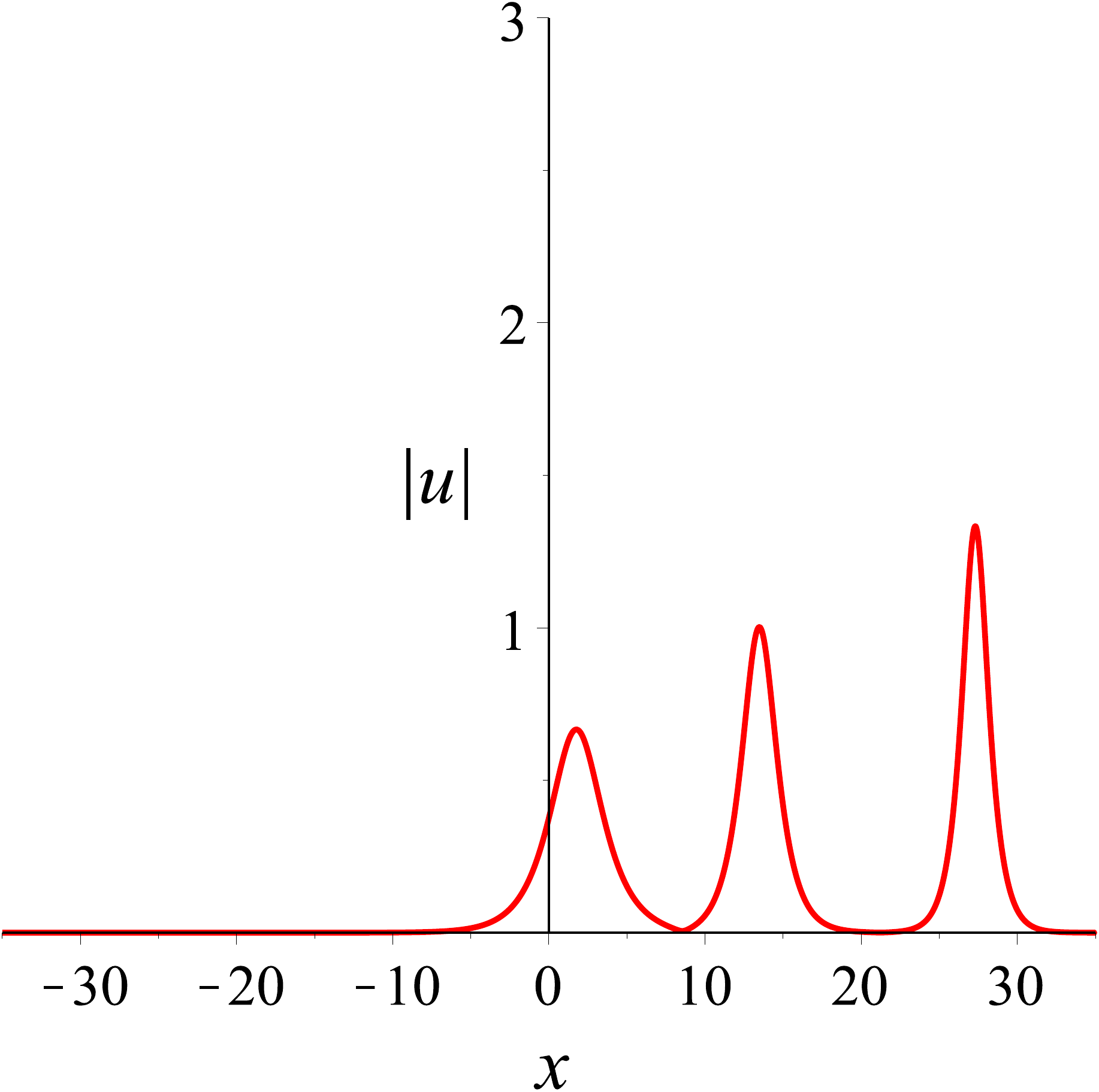}}}
\subfigure[]{\resizebox{0.31\hsize}{!}{\includegraphics*{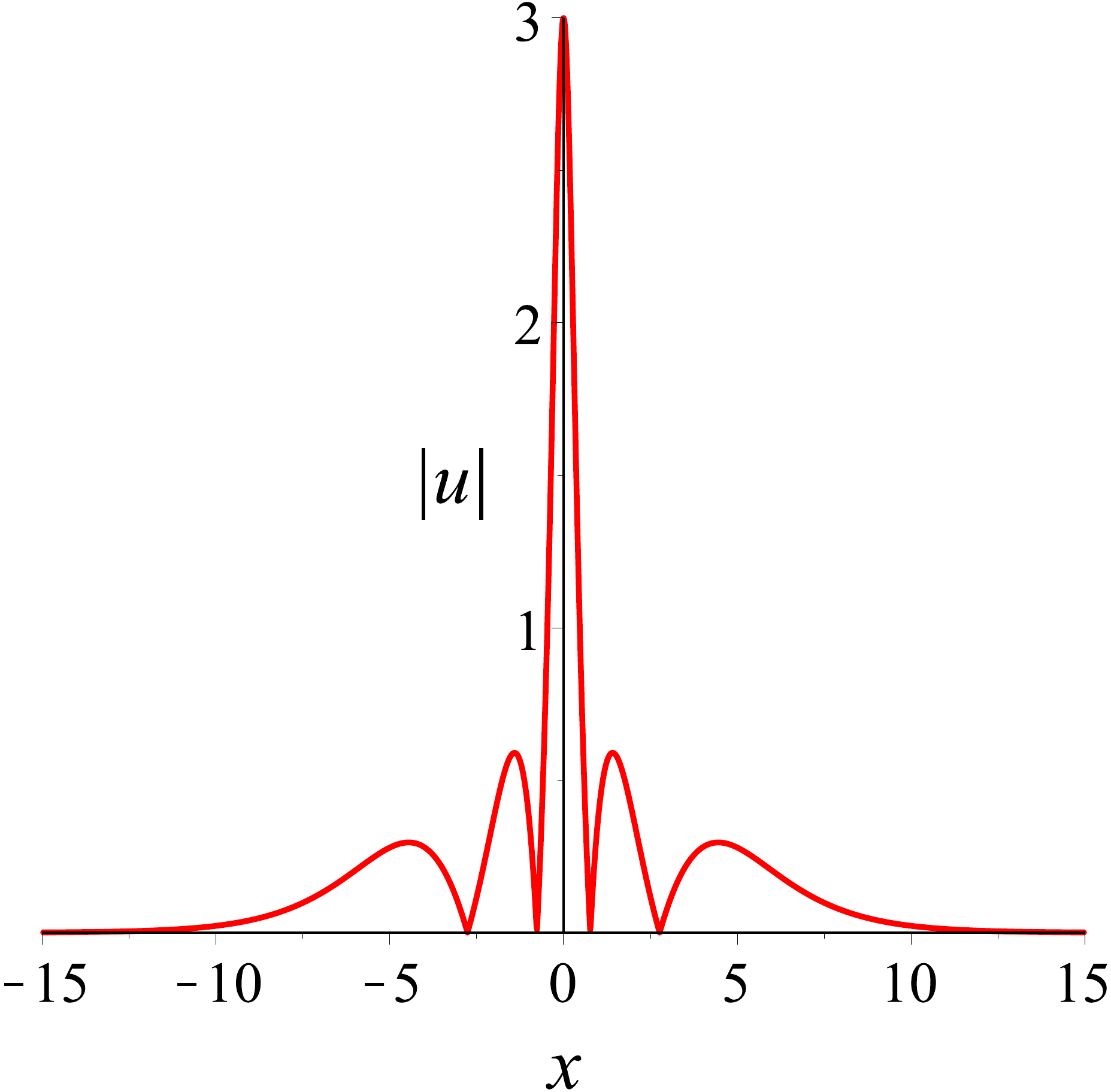}}}
\subfigure[]{\resizebox{0.31\hsize}{!}{\includegraphics*{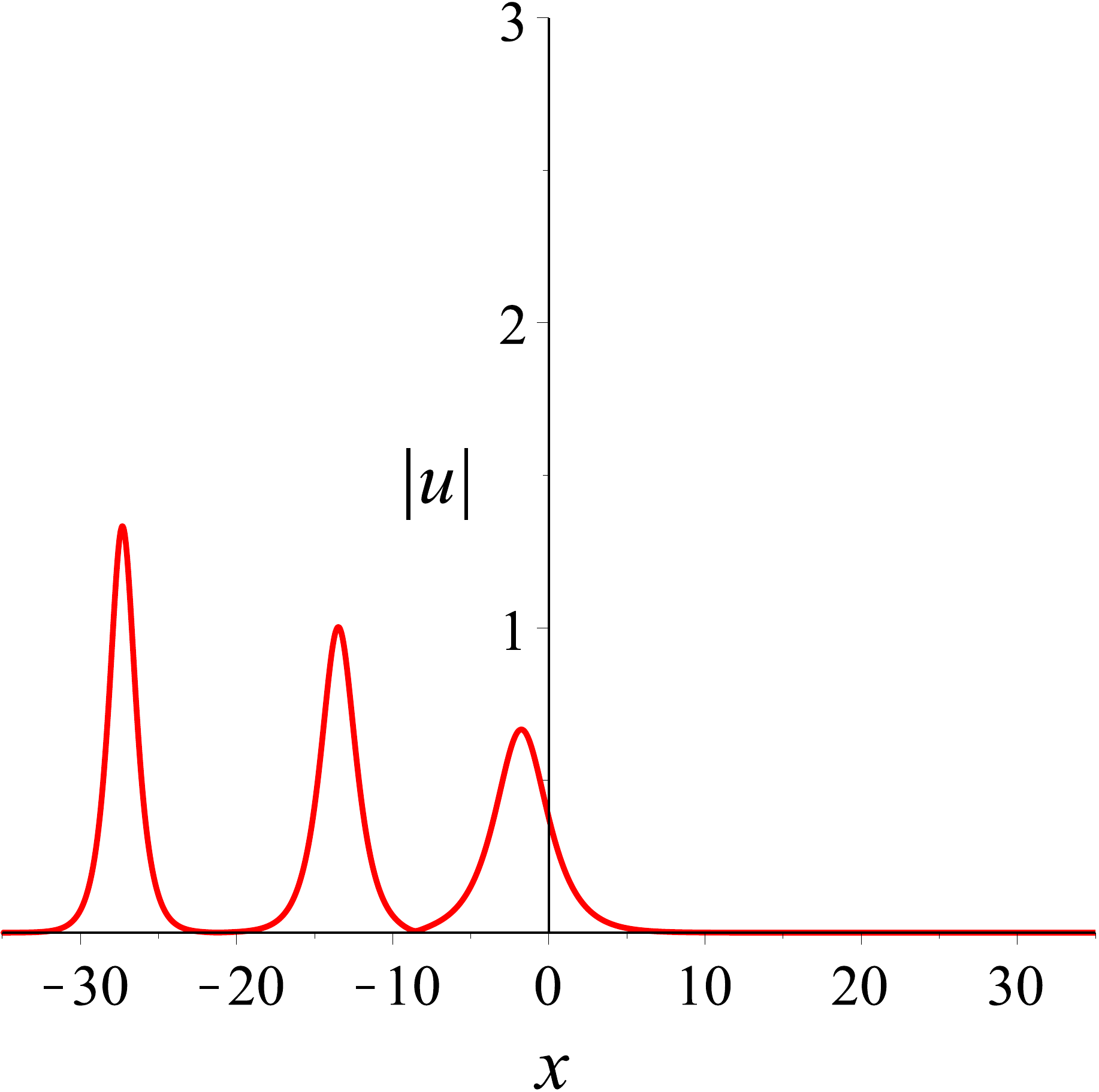}}}
\caption{Profiles of three-soliton solution (31) with $a_{1}=1,a_{2}=1,a_{3}=1,b_{1}=1,b_{2}=1,b_{3}=1,
\lambda_{1}=\frac{1}{10}+\frac{i}{2},
\lambda_{2}=\frac{1}{10}+\frac{2i}{3},\lambda_{3}=\frac{1}{10}+\frac{i}{3},
\epsilon =1,h=1$. (a) $x$-curve at $t=-18$; (b) $x$-curve at $t=0$; (c) $x$-curve at $t=18$.}
\end{center}
\end{figure}

\begin{figure}
\begin{center}
\subfigure[]{\resizebox{0.31\hsize}{!}{\includegraphics*{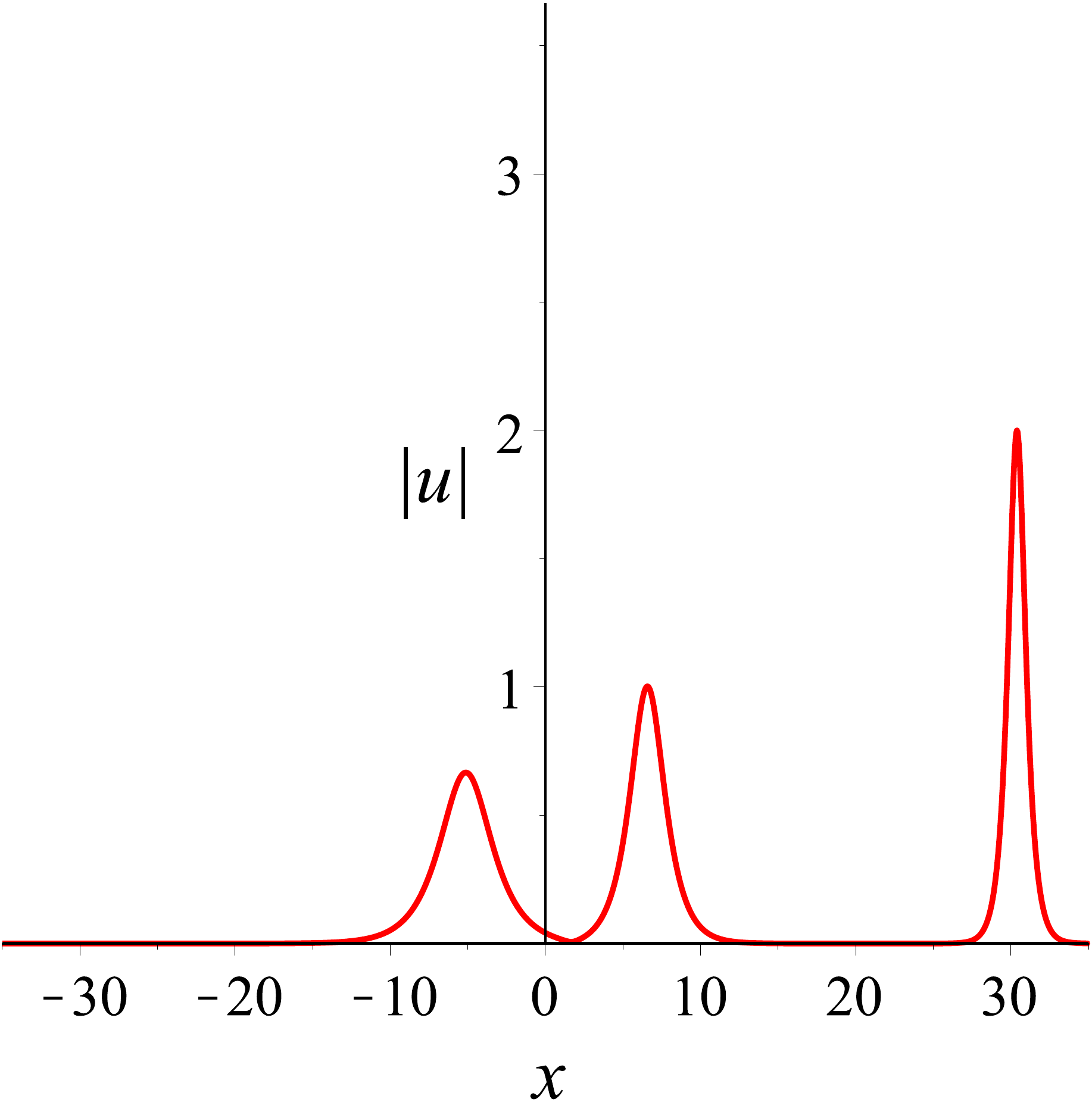}}}
\subfigure[]{\resizebox{0.31\hsize}{!}{\includegraphics*{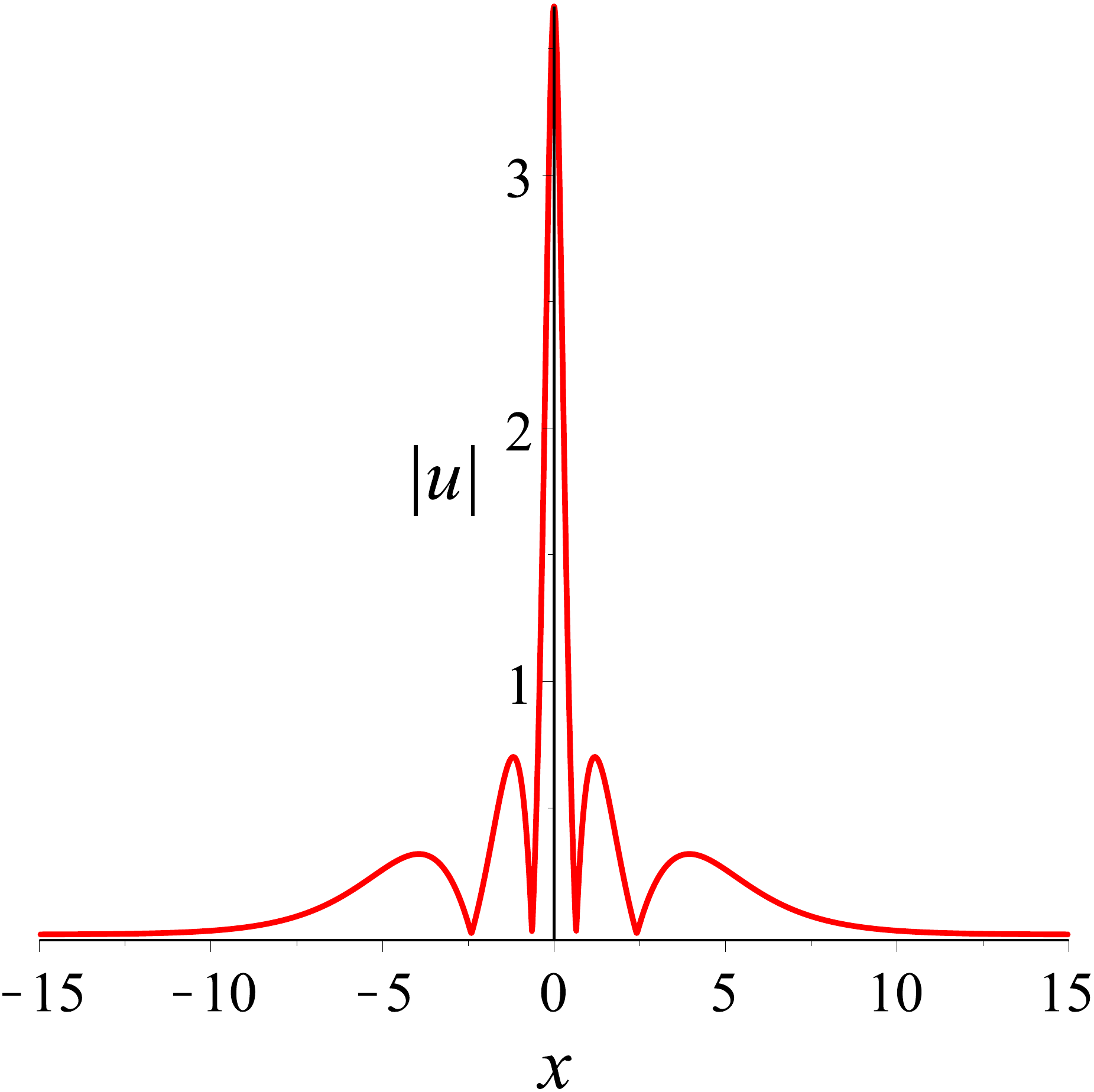}}}
\subfigure[]{\resizebox{0.31\hsize}{!}{\includegraphics*{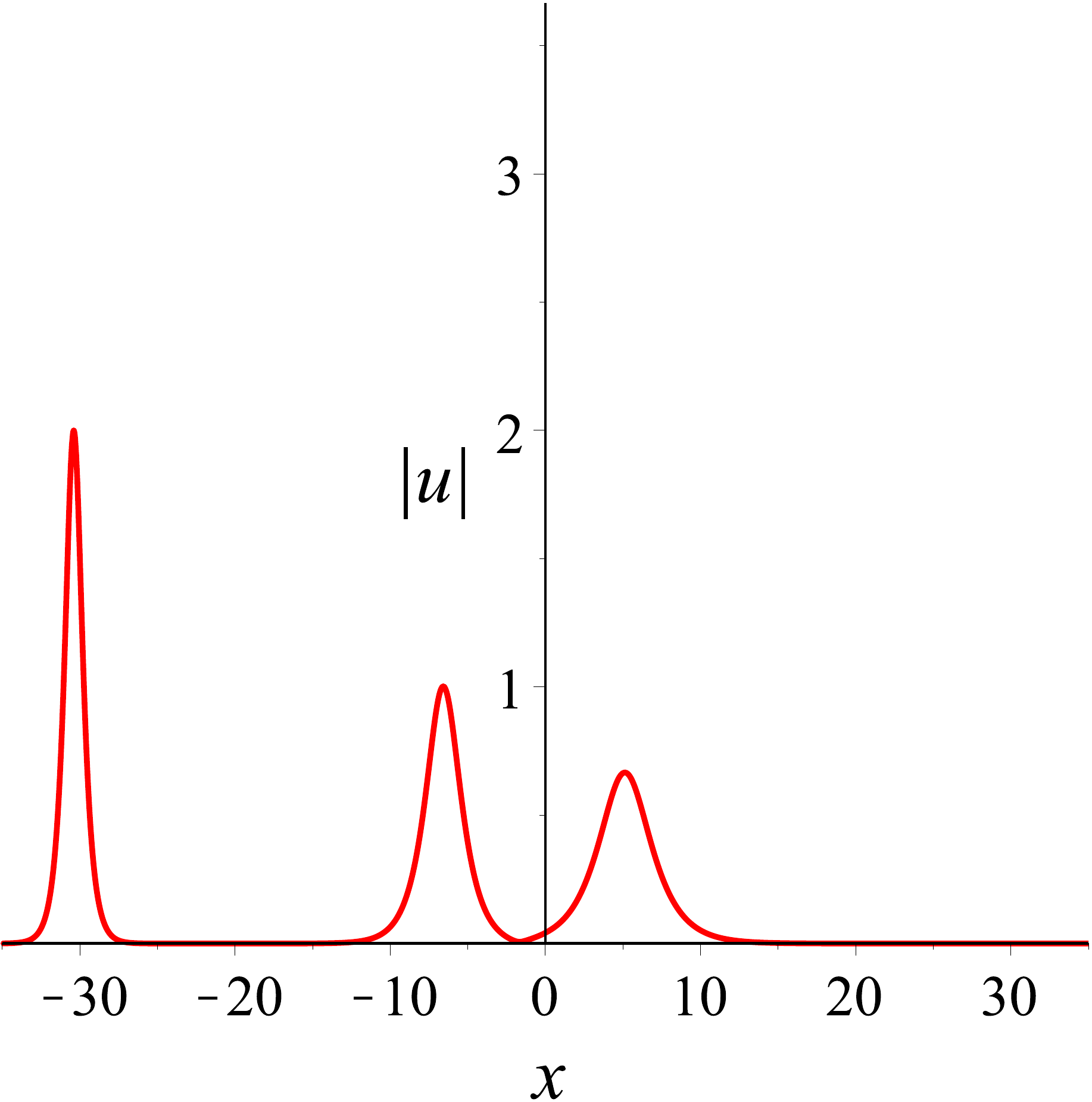}}}
\caption{Profiles of three-soliton solution (31) with $a_{1}=1,a_{2}=1,a_{3}=1,b_{1}=1,b_{2}=1,b_{3}=1,
\lambda_{1}=\frac{1}{10}+\frac{i}{2},
\lambda_{2}=\frac{1}{6}+\frac{i}{3},\lambda_{3}=\frac{1}{8}+i,
\epsilon=1,h=1$. (a) $x$-curve at $t=-8$; (b) $x$-curve at $t=0$; (c) $x$-curve at $t=8$.}
\end{center}
\end{figure}
\begin{figure}
\begin{center}
\subfigure[]{\resizebox{0.31\hsize}{!}{\includegraphics*{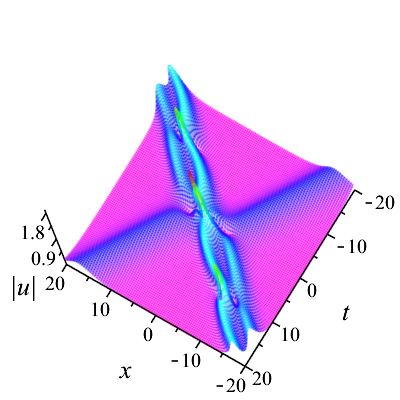}}}
\subfigure[]{\resizebox{0.31\hsize}{!}{\includegraphics*{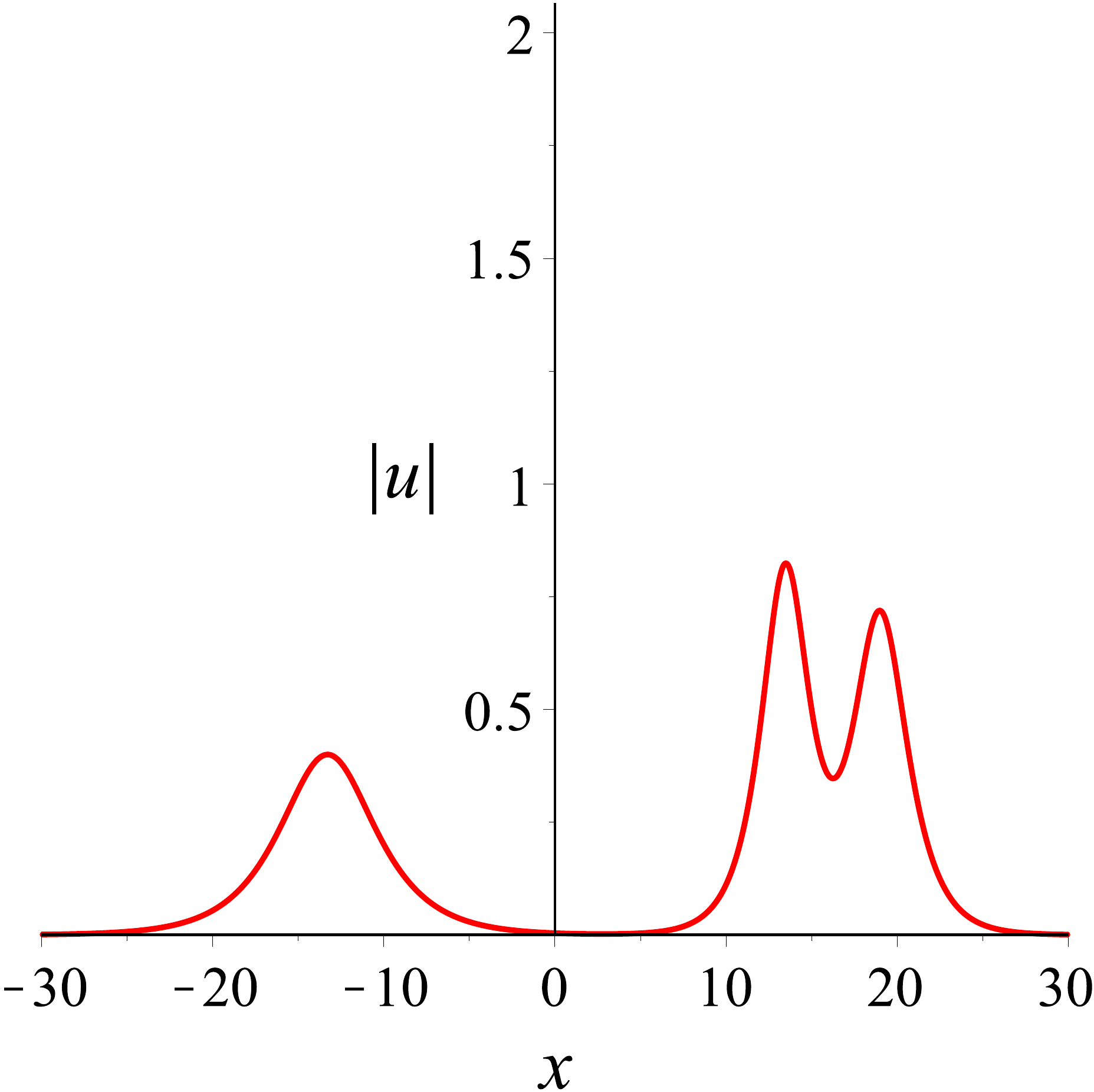}}}
\subfigure[]{\resizebox{0.31\hsize}{!}{\includegraphics*{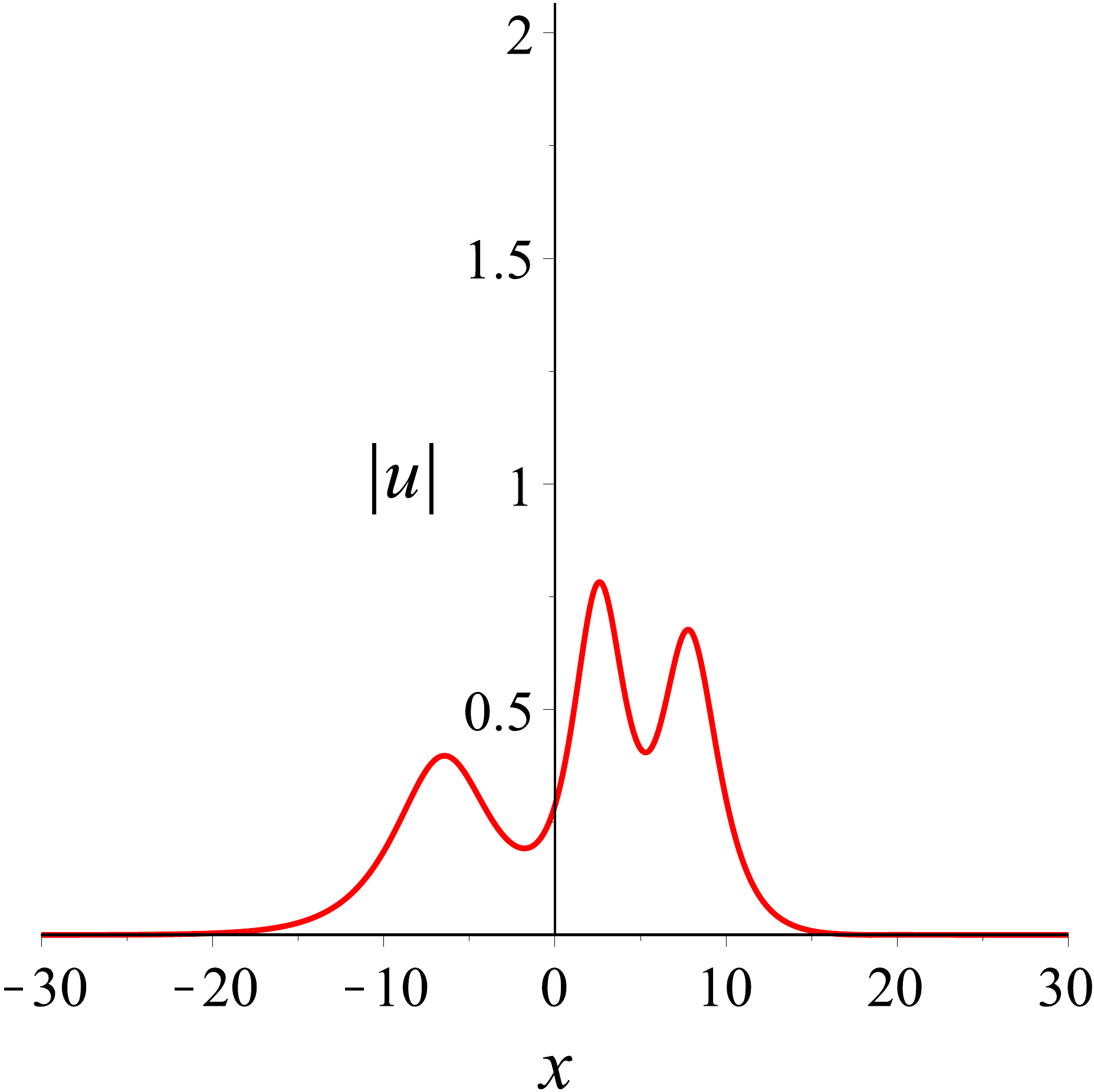}}}
\subfigure[]{\resizebox{0.31\hsize}{!}{\includegraphics*{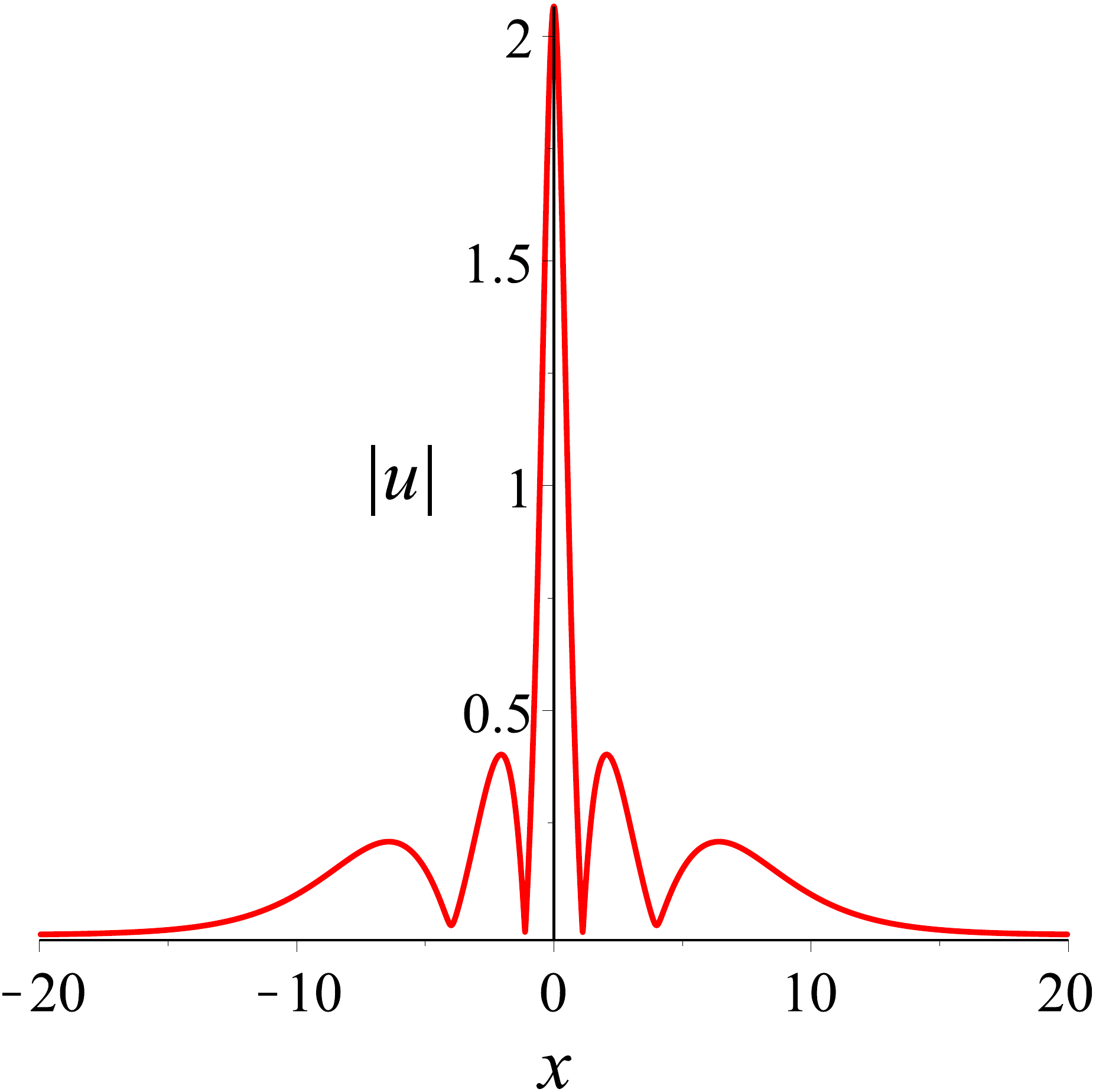}}}
\subfigure[]{\resizebox{0.31\hsize}{!}{\includegraphics*{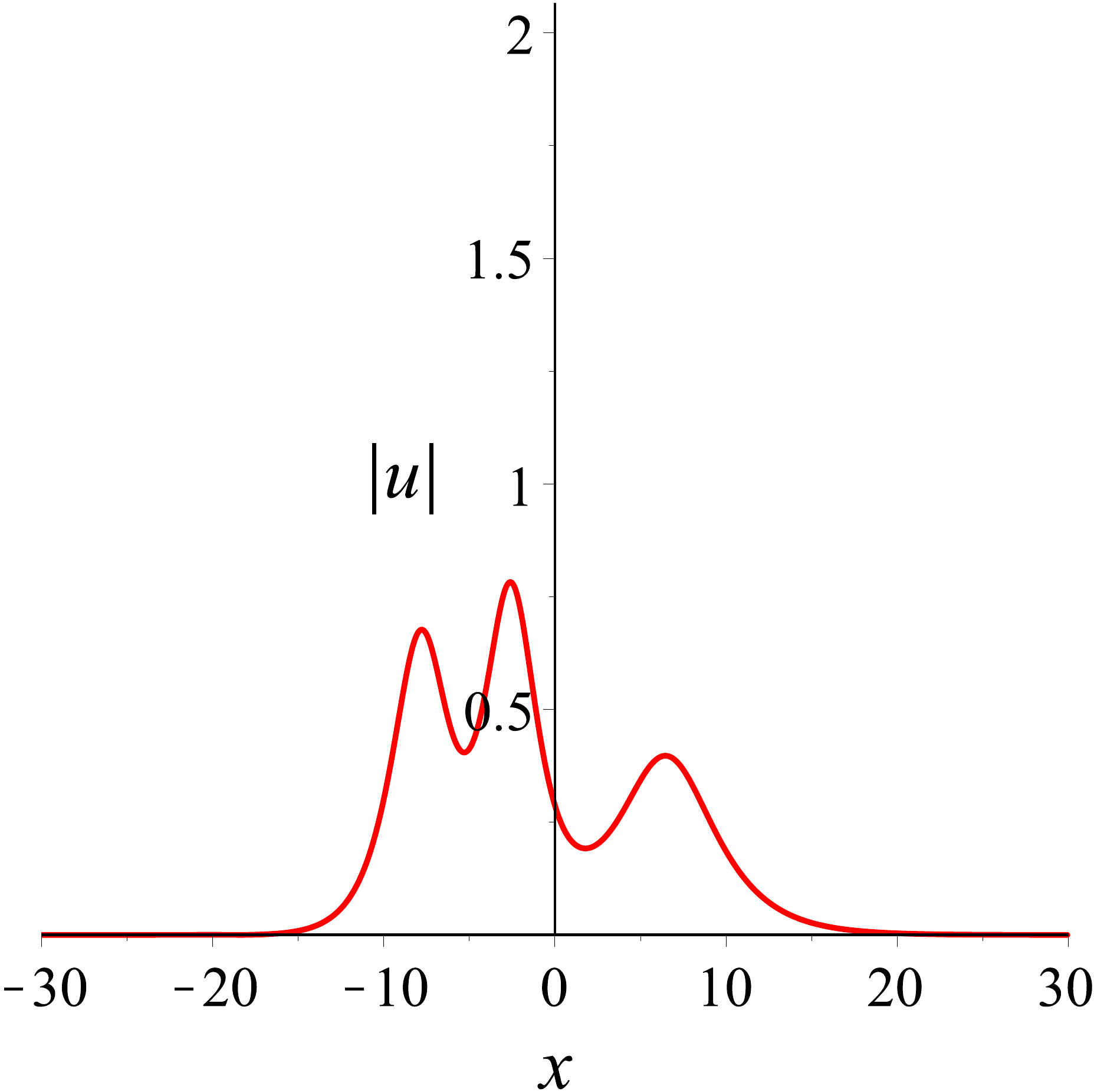}}}
\subfigure[]{\resizebox{0.31\hsize}{!}{\includegraphics*{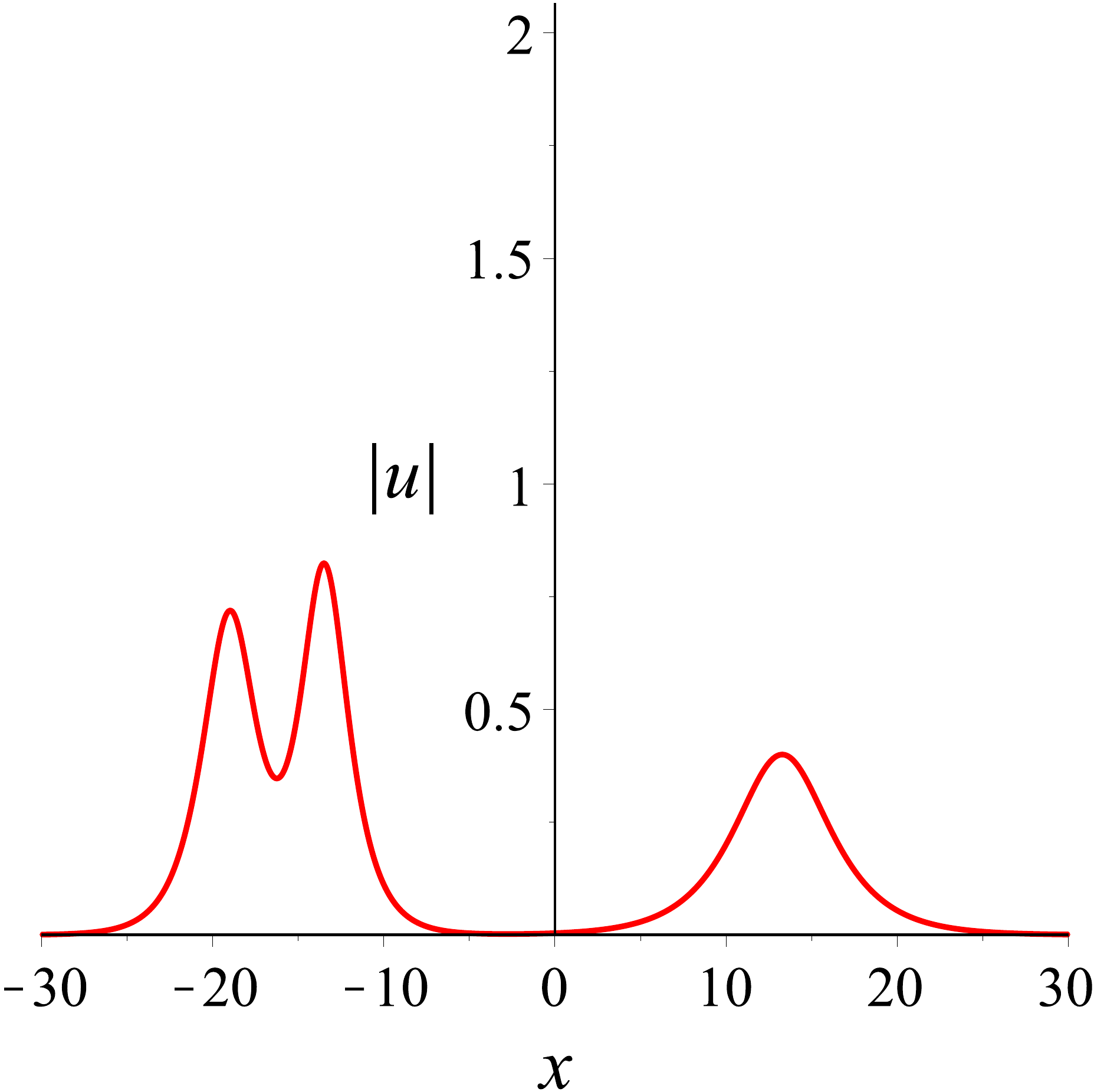}}}
\caption{Profiles of three-soliton solution (31) with $a_{1}=1,a_{2}=1,a_{3}=1,b_{1}=1,b_{2}=1,b_{3}=1,
\lambda_{1}=\frac{i}{2},
\lambda_{2}=\frac{1}{6}+\frac{i}{5},\lambda_{3}=\frac{i}{3},
\epsilon =1,h=1$. (a) 3D plot; (b) $x$-curve at $t=-15$; (c) $x$-curve at $t=-\frac{15}{4}$; (d) $x$-curve at $t=0$; (e) $x$-curve at $t=\frac{15}{4}$; (f) $x$-curve at $t=15$.}
\end{center}
\end{figure}
\begin{figure}
\begin{center}
\subfigure[]{\resizebox{0.31\hsize}{!}{\includegraphics*{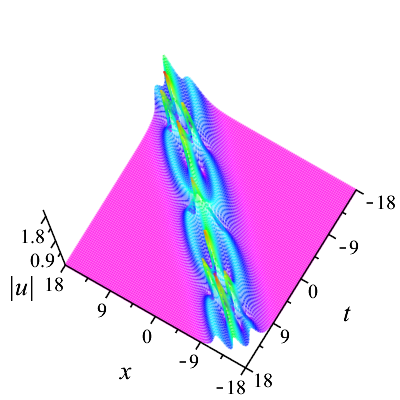}}}
\subfigure[]{\resizebox{0.31\hsize}{!}{\includegraphics*{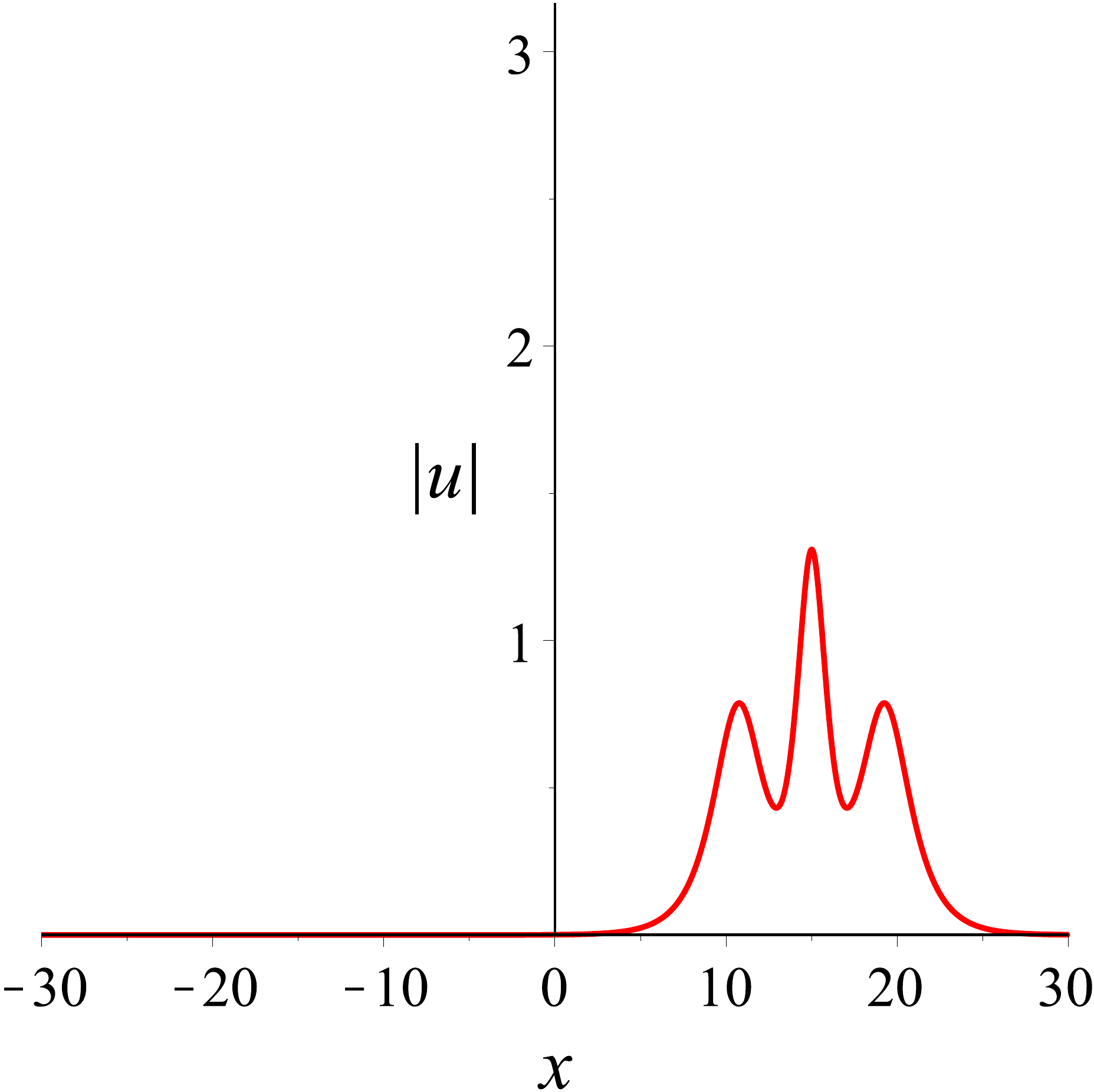}}}
\subfigure[]{\resizebox{0.31\hsize}{!}{\includegraphics*{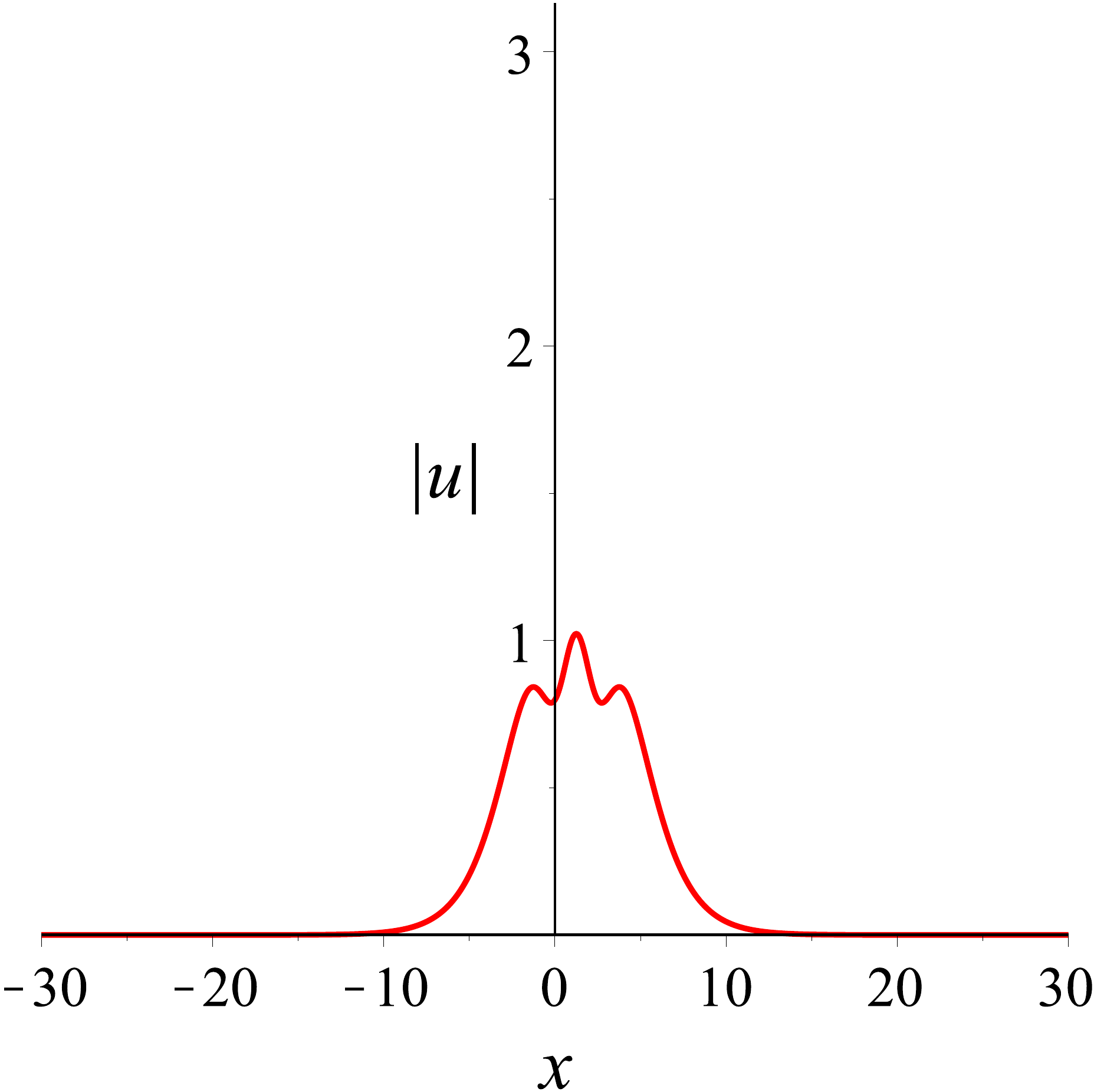}}}
\subfigure[]{\resizebox{0.31\hsize}{!}{\includegraphics*{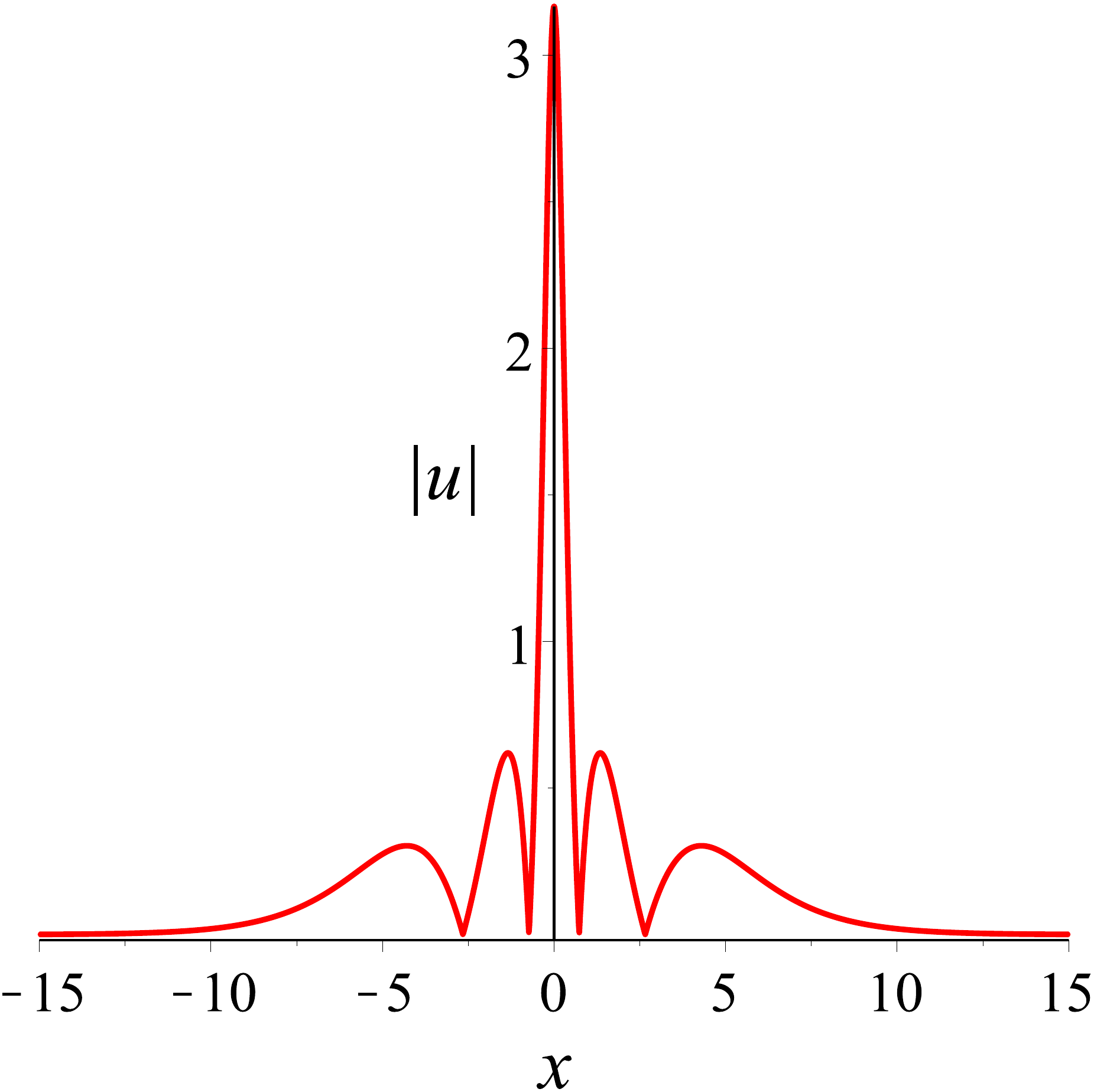}}}
\subfigure[]{\resizebox{0.31\hsize}{!}{\includegraphics*{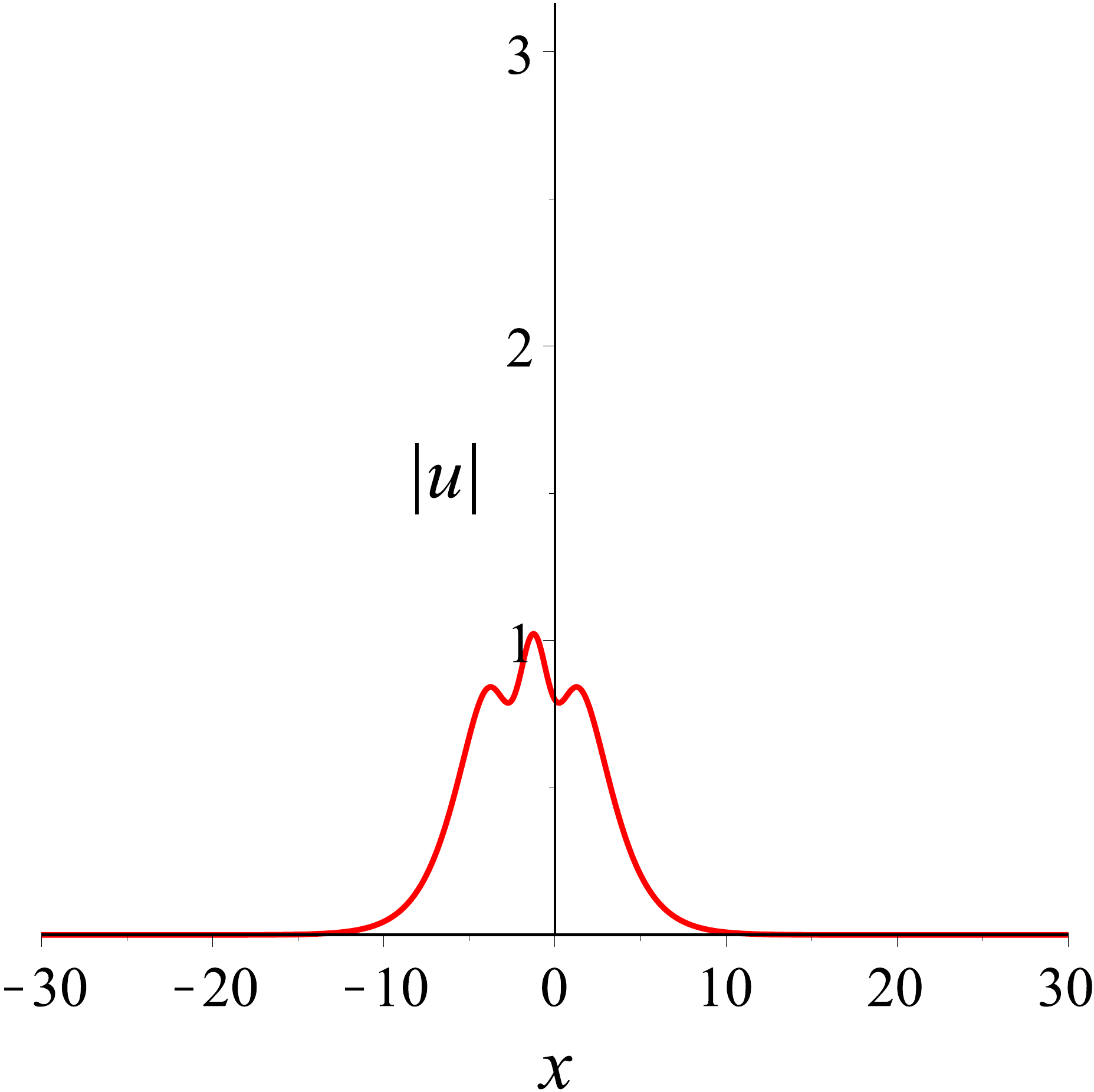}}}
\subfigure[]{\resizebox{0.31\hsize}{!}{\includegraphics*{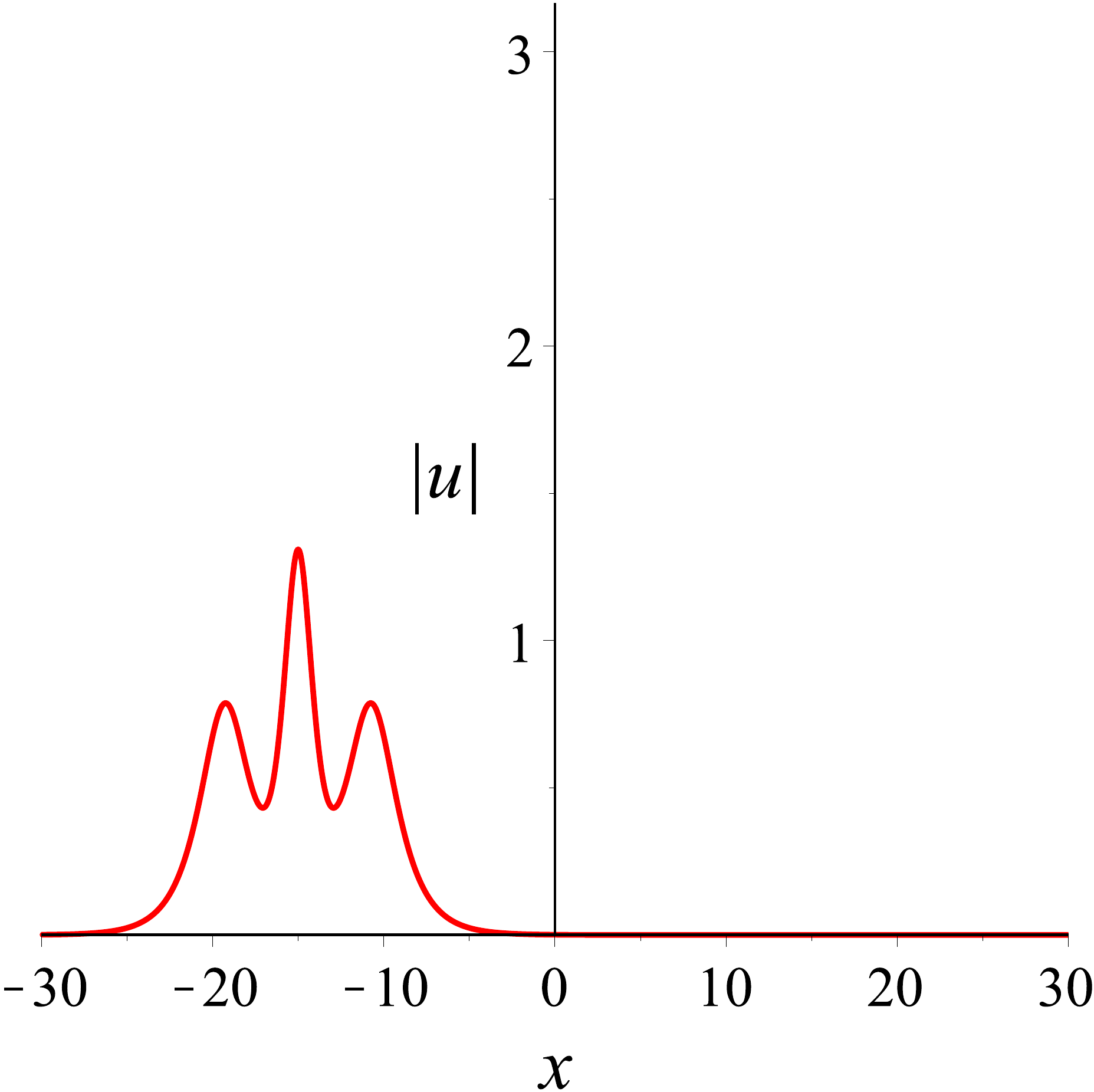}}}
\caption{Profiles of three-soliton solution (31) with $a_{1}=1,a_{2}=1,a_{3}=1,b_{1}=1,b_{2}=1,b_{3}=1,
\lambda_{1}=\frac{i}{2},\lambda_{2}=\frac{i}{3},\lambda_{3}=\frac{3i}{4},
\epsilon=1,h=1$. (a) 3D plot; (b) $x$-curve at $t=-15$; (c) $x$-curve at $t=-\frac{5}{4}$; (d) $x$-curve at $t=0$; (e) $x$-curve at $t=\frac{5}{4}$; (f) $x$-curve at $t=15$.}
\end{center}
\end{figure}

\section{Conclusion}
In this study, a generalized inhomogeneous higher-order nonlinear Schr\"odinger equation for
the Heisenberg ferromagnetic spin chain system in (1+1)-dimensions with the zero boundary condition was taken into account. A matrix Riemann-Hilbert problem was built, based on which multi-bright-soliton solutions to the examined equation were explored eventually. Moreover, the explicit forms of one-, two-, and three-bright-soliton solutions were given, and a few vivid plots were made to exhibit their spatial structures in three-dimensions and dynamical behaviors in two-dimensions after specifying the parameter values properly with the aid of Maple software.

\section*{Data availability}
Our manuscript has no associated data.

\section*{Conflict of interest}
The authors declare that they have no conflict of interest.

%\end{CJK*}  %% end the Chinese environment
%\end{document}  %%% end document
\newpage

\end{CJK*}  %% end the Chinese environment

\addcontentsline{toc}{chapter}{References}
\begin{thebibliography}{99}\footnotesize
\itemsep=-3pt plus.2pt minus.2pt   % set the reference line spacing

\bibitem{1} Enns, R.H., Jones, B.L., Miura, R.M., Rangnekar, S.S.: Nonlinear Phenomena in Physics and Biology. Springer, New York (1981)
\bibitem{2} Griffiths, G.W., Schiesser, W.E.: Linear and nonlinear waves. Scholarpedia {\bf4}, 4308 (2009)
\bibitem{3} Wen, X.K., Feng, R., Lin, J.H., Liu, W., Chen, F., Yang, Q.: Distorted light bullet in a tapered graded-index waveguide with PT symmetric potentials. Optik {\bf248}, 168092 (2021)
\bibitem{} Cao, Q.H., Dai, C.Q.: Symmetric and anti-symmetric solitons of the fractional second- and third-order nonlinear Schr\"odinger equation. Chin. Phys. Lett. {\bf38}, 090501 (2021)
\bibitem{} Wang, R.R., Wang, Y.Y., Dai, C.Q.: Influence of higher-order nonlinear effects on optical solitons of the complex Swift-Hohenberg model in the mode-locked fiber laser. Opt. Laser Tech. {\bf152}, 108103 (2022)
\bibitem{} Fang, J.J., Mou, D.S., Zhang, H.C., Wang, Y.Y.: Discrete fractional soliton dynamics of the fractional Ablowitz-Ladik model. Optik {\bf228}, 166186 (2021)
\bibitem{} Dai, C.Q., Wang, Y.Y.: Coupled spatial periodic waves and solitons in the photovoltaic photorefractive crystals. Nonlinear Dyn. {\bf102}, 1733--1741 (2020)
\bibitem{} Chen, Y.X.: Combined optical soliton solutions of a (1+1)-dimensional time fractional resonant cubic-quintic nonlinear Schr\"odinger equation in weakly nonlocal nonlinear media. Optik {\bf203}, 163898 (2020)
\bibitem{3} Wazwaz, A.M., El-Tantawy, S.A.: Solving the (3+1)-dimensional KP-Boussinesq and BKP-Boussinesq equations by the simplified Hirota's method. Nonlinear Dyn. {\bf88}, 3017--3021 (2017)
\bibitem{4} Zhang, S., Tian, C., Qian, W.Y.: Bilinearization and new multisoliton solutions for the (4+1)-dimensional Fokas equation. Pramana  {\bf86}, 1259--1267 (2016)
\bibitem{5} Yu, F.J., Feng, L.L., Li, L.: Darboux transformation for super-Schr\"odinger equation, super-Dirac equation and their exact solutions. Nonlinear Dyn. {\bf88}, 1257--1271 (2017)
\bibitem{6} Xu, T., Chen, Y.: Mixed interactions of localized waves in the three-component coupled derivative nonlinear Schr\"odinger equations. Nonlinear Dyn. {\bf92}, 2133--2142 (2018)
\bibitem{7} Ma, W.X.: Riemann-Hilbert problems and N-soliton solutions for a coupled mKdV system. J. Geom. Phys. {\bf132}, 45--54 (2018)
\bibitem{8} Wu, J.P.: Riemann-Hilbert approach of the Newell-type long-wave-short-wave equation via the temporal-part spectral analysis. Nonlinear Dyn. {\bf98}, 749--760 (2019)
\bibitem{9} Wu, J.P.: Integrability aspects and multi-soliton solutions of a new coupled Gerdjikov-Ivanov derivative nonlinear Schr\"odinger equation. Nonlinear Dyn. {\bf96}, 789--800 (2019)
\bibitem{10} Kumar, S., Kumar, A.: Lie symmetry reductions and group invariant solutions of (2+1)-dimensional modified Veronese web equation. Nonlinear Dyn. {\bf98}, 1891--1903 (2019)
\bibitem{11} Tanwar, D.V., Wazwaz, A.M.: Lie symmetries, optimal system and dynamics of exact solutions of (2+1)-dimensional KP-BBM equation. Phys. Scr. {\bf95}, 065220 (2020)
\bibitem{12} Wang, D.S., Yin, S.J., Tian, Y., Liu, Y.F.: Integrability and bright soliton solutions to the coupled nonlinear Schr\"odinger equation with higher-order effects. Appl. Math. Comput.  {\bf229}, 296--309 (2014)
\bibitem{13} Wang, D.S., Wang, X.L.: Long-time asymptotics and the bright $N$-soliton solutions of the Kundu-Eckhaus equation via the Riemann-Hilbert approach. Nonlinear Anal. Real World Appl. {\bf41}, 334--361 (2018)
\bibitem{14} Ma, W.X.: Riemann-Hilbert problems of a six-component fourth-order AKNS system and its soliton solutions. Comput. Appl. Math. {\bf37}, 6359--6375 (2018)
\bibitem{15} Ma, W.X.: Riemann-Hilbert problems and soliton solutions of a multicomponent mKdV system and its reduction. Math. Methods Appl. Sci. {\bf42}, 1099--1113 (2019)
\bibitem{16} Kang, Z.Z., Xia, T.C.: Construction of multi-soliton solutions of the $N$-coupled Hirota equations in an optical fiber. Chin. Phys. Lett. {\bf36}, 110201 (2019)
\bibitem{17} Kang, Z.Z., Xia, T.C., Ma, W.X.: Riemann-Hilbert method for multi-soliton solutions of a fifth-order nonlinear Schr\"odinger equation. Anal. Math. Phys. {\bf11}, 14 (2021)
\bibitem{18} Zhao, W.Z., Bai, Y.Q., Wu, K.: Generalized inhomogeneous Heisenberg ferromagnet model and generalized nonlinear Schr\"odinger equation. Phys. Lett. A {\bf352}, 64--68 (2006)
\bibitem{19} Sun, H., Shan, W.R., Tian, B., Wang, M., Tan, Z.: Analytic studies on a generalized inhomogeneous higher-order nonlinear Schr\"odinger equation for the Heisenberg ferromagnetic spin chain. Commun. Nonlinear Sci. Numer. Simulat. {\bf20}, 711--718 (2014)
\bibitem{20} Radha, R., Kumar, V.R.: Explode-decay solitons in the generalized inhomogeneous higher-order nonlinear Schr\"odinger equations. Z. Naturforsch. A  {\bf62}, 381--386 (2007)
\bibitem{18} Jia, H.X., Liu, Y.J., Wang, Y.N.: Rogue-wave interaction of a nonlinear Schr\"odinger model for the alpha helical protein. Z. Naturforsch. A {\bf71}, 27--32 (2016)
\bibitem{18} Zuo, D.W., Gao, Y.T., Xue, L., Sun, Y.H., Feng, Y.J.: Rogue-wave interaction for the Heisenberg ferromagnetism system. Phys. Scr. {\bf90}, 035201 (2015)
\bibitem{21} Wang, P., Qi, F.H., Yang, J.R.: Soliton solutions and conservation laws for an inhomogeneous fourth-order nonlinear Schr\"odinger equation. Comput. Math. Math. Phys. {\bf58}, 1856--1864 (2018)
\bibitem{15} Yang, J.K.: Nonlinear Waves in Integrable and Nonintegrable Systems. SIAM, Philadelphia (2010)
\end{thebibliography}
\end{document}